\documentclass[prb,superscriptaddress,twocolumn,showpacs,preprintnumbers,amsmath,amssymb,fixfloat]{revtex4-2}
\usepackage{graphicx}
\usepackage{amssymb}
\usepackage{dcolumn}
\usepackage{bm}
\usepackage{bbm}
\usepackage{float}
\usepackage{siunitx}
\usepackage{amsmath}

\usepackage{amsmath, amssymb}
\usepackage{bbold}
\usepackage{makecell}
\usepackage{tikz}
\usetikzlibrary{automata, positioning}
\usetikzlibrary{arrows}
\usepackage{multirow}
\usepackage{mathtools}
\def\block(#1,#2)#3{\multicolumn{#2}{c}{\multirow{#1}{*}{$ #3 $}}}

\begin{document}

\title{Making every photon count: A quantum polyspectra approach to the dynamics of  blinking quantum emitters at low photon rates without binning}
\author{M. Sifft}
\address{Ruhr University Bochum, Faculty of Physics and Astronomy, Experimental Physics VI (AG), Universit\"atsstra{\ss}e 150, D-44780 Bochum, Germany}
\author{A. Kurzmann}
\address{Faculty of Physics and CENIDE, University of Duisburg-Essen, Lotharstra{\ss}e 1, 47057 Duisburg, Germany}
\author{J. Kerski}
\address{Faculty of Physics and CENIDE, University of Duisburg-Essen, Lotharstra{\ss}e 1, 47057 Duisburg, Germany}
\author{R. Schott}
\address{Lehrstuhl f\"ur Angewandte Festk\"orperphysik, Ruhr-Universit\"at Bochum, Universit\"atsstra{\ss}e 150, D-44780 Bochum, Germany}
\author{A. Ludwig}
\address{Lehrstuhl  f\"ur Angewandte Festk\"orperphysik, Ruhr-Universit\"at Bochum, Universit\"atsstra{\ss}e 150, D-44780 Bochum, Germany}
\author{A. D. Wieck}
\address{Lehrstuhl  f\"ur Angewandte Festk\"orperphysik, Ruhr-Universit\"at Bochum, Universit\"atsstra{\ss}e 150, D-44780 Bochum, Germany}
\author{A. Lorke}
\address{Faculty of Physics and CENIDE, University of Duisburg-Essen, Lotharstra{\ss}e 1, 47057 Duisburg, Germany}
\author{M. Geller}
\address{Faculty of Physics and CENIDE, University of Duisburg-Essen, Lotharstra{\ss}e 1, 47057 Duisburg, Germany}
\author{D. H\"agele}
\address{Ruhr University Bochum, Faculty of Physics and Astronomy, Experimental Physics VI (AG), Universit\"atsstra{\ss}e 150, D-44780 Bochum, Germany}

\date{\today}
\begin{abstract} 
The blinking statistics of quantum emitters and their corresponding Markov models play an important role in high resolution microscopy of biological samples as well as in nano-optoelectronics and many other fields of science and engineering. Current methods for analyzing the blinking statistics like the full counting statistics or the Viterbi algorithm break down for low photon rates. 
We present an evaluation scheme that eliminates the need for both a minimum photon flux and the usual binning of photon events which limits the measurement bandwidth. Our approach is based on higher order spectra of the measurement record which we model within the recently introduced method of quantum polyspectra from the theory of continuous quantum measurements. By virtue of this approach we can determine on- and off-switching rates of a semiconductor quantum dot at light levels 1000 times lower than in a standard experiment and 20 times lower than achieved with a scheme from full counting statistics. Thus a very powerful high-bandwidth approach to the parameter learning task of single photon hidden Markov models has been established with applications in many fields of science.  
\end{abstract}

 \maketitle

\section{Introduction}
\begin{figure}[t]
	\centering
	\includegraphics[width=8.6cm]{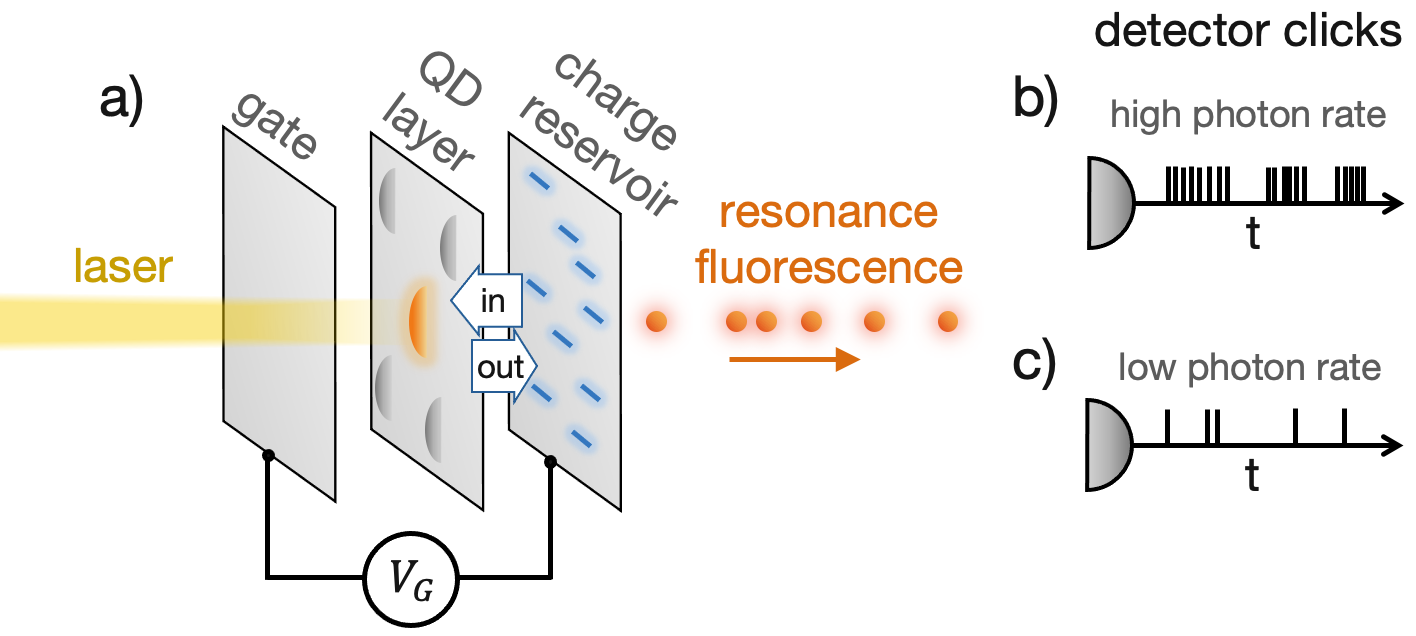}
	\caption{(a) A single semiconductor quantum dot coupled to a charge reservoir switches incoherently between a bright (charged) and dark (neutral) state due to carrier tunneling. (b) Single photons of the resonance fluorescence appear as stochastic peaks in the measurement record. The bright and dark state dynamics is directly visible only at high photon rates. (c) Access to charging statistics at low photon rates via higher order spectra of the detector output
is demonstrated in this work.}
	\label{figure0}   
\end{figure}

The light emission of so-called quantum emitters is dominated by the specific quantum mechanics of the emitting system which often can described by just a few quantum states. Their single photon emission clearly distinguishes them from thermal light sources - like light bulbs - or coherent sources - like lasers - via their photon statistics \cite{stevensREVIEW2013}.
Quantum emitters play an important role in many fields of science. In biology, chromophores behave as quantum emitters in fluorescence microscopy. They are used to label proteins for the investigation of protein dynamics. Proteins switching between different configurations can be characterized via the changing brightness of chromophores which is strongly influenced by Förster resonance transfer (FRET) \cite{AndrecPCA2003,mckinneyBIOPHYSJ2006,JagerChemPhysChem2009,gotz2022blind}. In physics, semiconductor quantum dots are quantum emitters that are currently investigated as single photon sources in quantum information devices \cite{senellartNATURE2017, GschreyNatComm2015, HeindelNJP2012, MuellerNatPho2014} or as probes of charging dynamics \cite{kurzmannPRL2019, VamivakasNATURE2010, KuhlmannNATPHY2013}.
Figure \ref{figure0}(a) displays the schematics of a semiconductor quantum dot whose incoherent charging dynamics is investigated by time-resolved detection of single photons. In case of high photon rates the bright and dark states of the blinking quantum dot can be clearly distinguished after binning of single photon clicks over a finite time interval [see, e.g., Fig. \ref{figure0}(b)]. The resulting measurement trace exhibits random telegraph noise whose statistics coincides with the statistics of the charging dynamics. The theory of hidden Markov models (HMM) \cite{mckinneyBIOPHYSJ2006}, the full counting statistics (FCS) of transport theory \cite{levitovJMP1996}, and also quantum polyspectra \cite{sifftPRR2021} provide methods for fitting system parameters to simple models that capture the essence of the stochastic dynamics. The limitations of binning have recently been discussed in the context of the FCS by Kerski {\it et al.} \cite{kerskiSR2023}. They found that the measurable bandwidths can in their scheme only be increased by a higher photon rate.   

Therefore, the following challenging question appears in the case of low photon rates [Fig. \ref{figure0}(c)]: Is it fundamentally possible to infer the blinking statistics from photon click events if the photon rates drop below the switching rate of the quantum dot, i.e. if sometimes no photon is detected even during the quantum dot being in the bright state? Obviously, no sensible telegraph signal can be obtained from binning single click events [see Fig. \ref{figure0}(c) and Fig. \ref{sample_poly}(a)]. 
In the following, we will give a positive answer to the question by modeling the system of quantum dot {\it and} single photon detector within the theory of continuous quantum measurements. Recent advances in the theory established a connection between system properties (in our case transition rates) and the continuous measurement of an observable (the presence of the photon) in terms of its higher order spectra \cite{hagelePRB2018,hagelePRB2020E,sifftPRR2021,sifftPRA2023,meinelNATURECOMM2022}. Such quantum polyspectra provide here the key for solving the problem.
The present work extends our previous scheme for analyzing quantum dot dynamics from the continuous measurement regime to the single photon regime establishing a very general alternative to the FCS \cite{sifftPRR2021,kleinherbersPRL2022,landiARXIV2023}. 

The paper is organized as follows. 
Sec. II shows that higher order spectra of single photon data  keep many features even for decreasing photon rates nurturing the hope that system statistics can be recovered even at very low photon rates.
In Sec. III we shortly review how continuous Markov models are used to model random telegraph noise and what approaches have been used in the past to recover transition rates from experimental data. We show how a Markov model for telegraph noise can be transformed into a modified Markov model which exhibits single photon events.
We explain how theoretical polyspectra of this model can be calculated by mapping it onto a quantum model which then can be treated within the very general framework of so-called quantum polyspectra. 
In Sec. IV we use real-world data to test our new approach and find that transition rates can be extracted with the polyspectral approach even in cases of extremely low photon rates. A relation of our work to the theory of Cox-processes from mathematical statistics is discussed in Sec. \ref{sec:discussion} \cite{coxJRSSSB1955}.
\section{Higher order spectra of single photon data} 
\begin{figure*}[t]
	\centering
	\includegraphics[width=17.6cm]{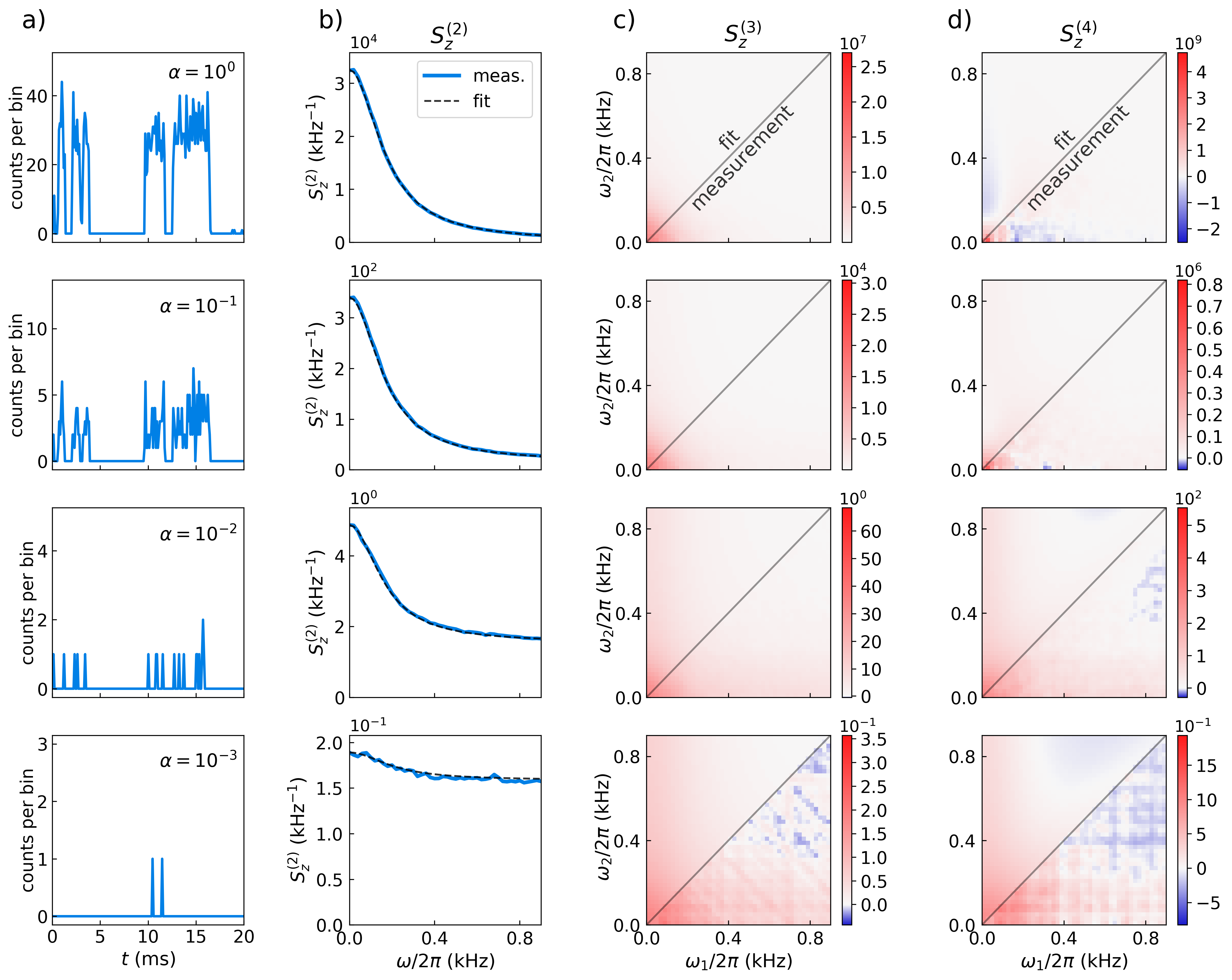}
	\caption{(a) Samples of the measurement record for photon fractions of $\alpha = 1,\,10^{-1},\,10^{-2},\,10^{-3}$. Binning into $100$~$\mu$s intervals was used for visualizing single photon events. (b-d) Experimental polyspectra up to fourth order obtained without binning (App.  \ref{app:NumericalSpectra}) and their analytic counterparts that followed from a fitting procedure to quantum polyspectra of a Markov model. Spectra $S_z^{(3)}$ and $S_z^{(4)}$ are given in units of kHz$^{-2}$ and kHz$^{-3}$, respectively.
	The overall backgrounds found in $S_z^{(3)}$ and $S_z^{(4)}$ were subtracted in the graphs for a better visibility of the spectral structure on top.}
	\label{sample_poly}   
\end{figure*}

In this work we aim to analyze single photon data measured by Kurzmann {\it et al.} on a single InAs quantum dot embedded in a p-i-n diode matrix \cite{kurzmannPRL2019, KurzmannSupp2019}.
Figure \ref{sample_poly}(a) shows a small part of a measurement trace of six minute duration. Single photon clicks with time stamps $t_j$ were binned into \SI{100}{\micro s} intervals here only for display. However, no binning is required for calculating the polyspectra. The number of clicks was artificially reduced to a photon fraction of $\alpha = 10^{-1},\, 10^{-2}$, and $10^{-3}$  by randomly deleting any single click-event with a probability of $1-\alpha$ from the original data set (see lower rows) using a standard random number generator.
Clearly, the telegraph noise behavior visible in the first row gets completely compromised at reduced photon rates raising the question if the underlying on-off behavior of the quantum emitter can still be inferred from such data. 
Figure \ref{sample_poly}(b) shows power spectra $S_z^{(2)}(\omega)$ calculated from the measurement trace $z(t)$ without binning (see App. \ref{app:NumericalSpectra}). While a spectrally flat background increases as $\alpha$ decreases, we interestingly find that a peak at zero frequency prevails independently from the photon fraction $\alpha$. Clearly, some information about the measured system survives even for strong photon loss. 
It is known that the usual power spectrum $S_z^{(2)}(\omega)$ reveals only the sum of transition rates in case of a two-state model while their separation requires a higher order spectrum \cite{sifftPRR2021}. Specifically, $S_z^{(3)}$ contains information about the difference between the transition rates, while $S_z^{(4)}$ depends on higher-order polynomials of the transition rates.
We are therefore interested also in higher order spectra $S_z^{(3)}$ and $S_z^{(4)}$ [Figs.  \ref{sample_poly}(c) and  \ref{sample_poly}(d)] \cite{hagelePRB2018,sifftPRR2021,sifftPRA2023}. 
Similarly to $S_z^{(2)}$, the higher order spectra show an increasing background for decreasing photon fraction $\alpha$. 
Nevertheless, we will show in Sec. \ref{sec:Markov} and Sec. \ref{sec:analysis} that a simultaneous fitting of all spectra with
model spectra will correctly recover the on-off transition rates of the quantum emitter. The model spectra obtained from the fitting procedure a shown in Fig \ref{sample_poly} along with the measured spectra.

Brillinger’s polyspectra generalize the usual power spectrum $S_z^{(2)}(\omega) \propto \langle z(\omega) z^*(\omega) \rangle$ of a stochastic process $z(t)$ to spectra that are of higher orders of $z(\omega)$ with
$z(\omega) = \int z(t) e^{i \omega t}\,dt$ being the Fourier transform of $z(t)$ \cite{brillingerAMS1965}. 
The definition of polyspectra
\begin{eqnarray}
	2\pi \delta(\omega_1+...+\omega_n)&S_z^{(n)}&(\omega_1,...,\omega_{n-1})\nonumber\\ 
	&=& C_n(z(\omega_1),..., z(\omega_n)), \label{eq:defPolyspectra}
\end{eqnarray} 
is based on the $n$th-order cumulant $C_n$, where
\begin{eqnarray}
  C_2(x,y) & = & \langle x y \rangle - \langle x \rangle \langle  y \rangle \nonumber \\
 C_3(x,y,z) & = & \langle (x - \langle x \rangle) ( y - \langle y \rangle)  (z - \langle z \rangle) \rangle. 
\end{eqnarray}
The fourth order cumulant can, e.g., be found in Refs. 
 \cite{gardinerBOOK2009,hagelePRB2018}.
The bi\-spectrum $S_z^{(3)}$ is related to $\langle z(\omega_1) z(\omega_2) z^*(\omega_1 + \omega_2)\rangle$ and exhibits a non-vanishing imaginary part in the case of broken time-inversion symmetry of $z(t)$. A cut through the trispectrum $S_z^{(4)}(\omega_1,\omega_2,-\omega_1)$ is used in this article. This two-dimensional spectrum is related to 
 $\langle z(\omega_1) z^*(\omega_1) z(\omega_2) z^*(\omega_2)\rangle – \langle z(\omega_1) z^*(\omega_1) \rangle \langle z(\omega_2) z^*(\omega_2)\rangle$ and can be interpreted as a correlation spectrum of intensity fluctuation of $z(t)$ \cite{starosielecRSI2010}. A recipe for the estimation of polyspectra from data can be found in App. \ref{app:NumericalSpectra}. Its implementation is included in our SignalSnap software library and was used here for calculating experimental polyspectra \cite{sifftSIGNALSNAP2022}.
\section{Continuous Markov Models}
\label{sec:Markov}
In this section, Markov models are used to model single photon measurement traces. Analytical expressions for polyspectra of such traces are found within the framework of quantum polyspectra \cite{hagelePRB2018}. 

The theory of quantum polyspectra has originally been developed for calculating polyspectra of continuous measurements on general open quantum systems. Here we apply the framework to continuous Markov systems which can be viewed as quantum systems showing only incoherent dynamics without any quantum coherence. Keeping the full theory allows for an easy future adaption of our method to quantum systems which exhibit coherent dynamics of e.g. a precessing electron spin \cite{sifftPRA2023}. A specific theory of polyspectra for continuous Markov models would, however, be numerically less demanding and is currently being developed and implemented in our group. 
Polyspectra play a key role in Sec. \ref{sec:analysis}, when they are fitted to experimental polyspectra for eventually finding system parameters with which the statistics of measurement traces can be matched. 

\begin{figure}[t]
	\centering
	\includegraphics[width=8.6cm]{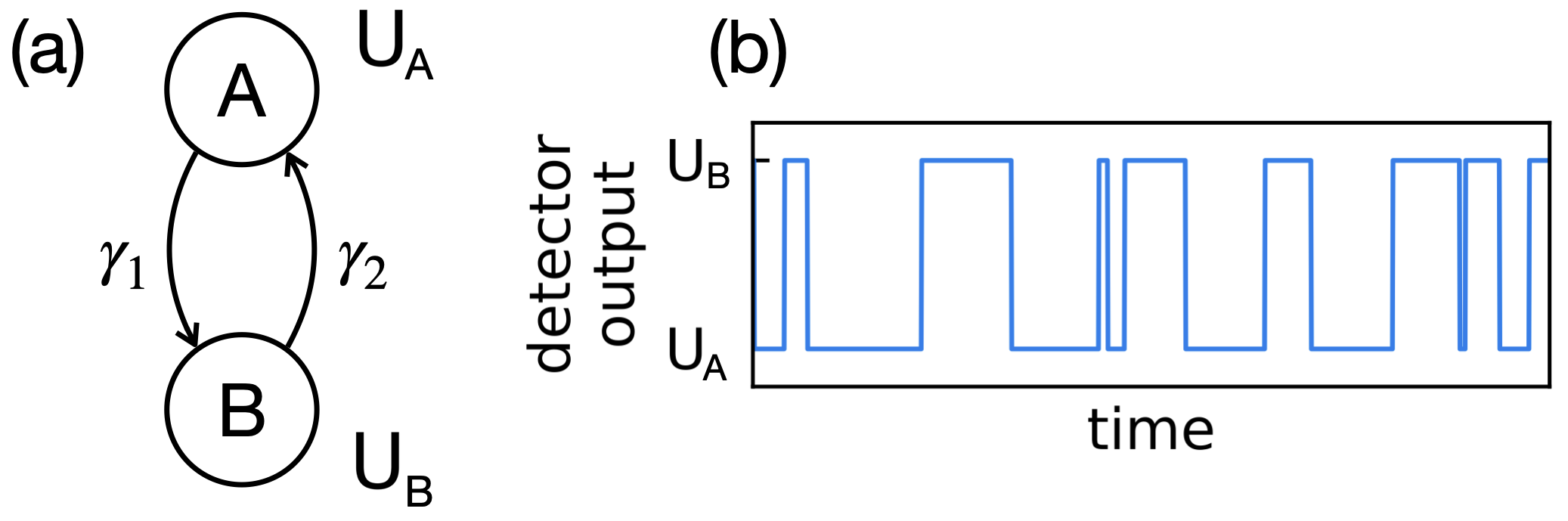}
	\caption{Continuous Markov-model with states A and B and random telegraph noise with corresponding signal levels $U_{\rm A}$ and $U_{\rm B}$.}
	\label{figure_Markov}   
\end{figure}

Markov models find application in science to describe systems that stochastically switch between a  number of discrete states. The switching rates towards a new state depend only on the actual state. Therefore, the future behavior of Markov models does not depend on its past making them 
so-called memory-less models. Figure \ref{figure_Markov}(a) depicts a simple system that switches between two states A and B with transition rates $\gamma_1$ and $\gamma_2$. The observation of a corresponding real-world system would give rise to a time-dependent signal $U(t)$ switching between output-signals $U_{\rm a}$ and $U_{\rm b}$ [see Fig. \ref{figure_Markov}(b)]. The challenge posed by such a measured trace is to find the corresponding hidden Markov model (HMM). In biology, a number of related methods for finding model parameters are in use like the Viterbi-algorithm, the forward-algorithm, and the Baum-Welch-algorithm, to name but a few \cite{forneyIEEE1973,rabinerIEEE1986,mckinneyBIOPHYSJ2006}. They have in common that model parameters are varied with the aim to increase the probability of reproducing the measured trace $U(t)$. While the probability of obtaining a specific measurement outcome is of course extremely small, this quantity is still very valuable for judging the quality of a Markov Model. The algorithms differ mostly in their ability to cope with measurement noise. A common drawback of all the algorithms is that the full measurement trace must be stored and evaluated every time a new set of model parameters is investigated. This limits the applicability of such algorithms to short measurement traces.
This problem becomes even larger for HMM methods where, instead of binned photon signals, single photon events are considered \cite{JagerChemPhysChem2009}.  
The polyspectra approach presented here does overcome this limitation. Deliberately long measurement traces are evaluated only once in terms of their polyspectra which require much less storage memory than the initial measurement trace. The polyspectra rather than the full measurement trace enter the subsequent fitting routines for reconstructing the Markov model.
In the field of quantum electronics, a pure Markov-approach is often sufficient to describe the observed system dynamics and the statistics of corresponding measurement records. However, a completely different tool-set for analyzing data has been developed in this field. The so-called full counting statistics (FCS) relies on the identification of quantum jumps in the measurement trace and treats quantities like the probability $P(N,t)$ of $N$ electrons leaving a quantum dot in time $t$ \cite{levitovJMP1996,bagretsPRB2003,ubbelohdeNATCOMM2012,emaryPRB2007}. Recently, factorial cumulants based on $P$ have been shown to be useful for analyzing compromised binned data where small photon numbers can lead to randomly appearing wrong counts of jump events. Kleinherbers {\it et al.} evaluated the same data as in this article at a photon fraction of $\alpha = 2 \times 10^{-2}$ \cite{kleinherbersPRL2022} based on factorial cumulants of the FCS \cite{kamblyPRB2011,stegmannPRB2015}. Here, we push the limit to $\alpha = 10^{-3}$  without even compromising temporal resolution by binning.

\subsection{Markov-description of single photon detection}
The Markov model from above approximates sufficiently the appearance of a measurement trace that exhibits distinct levels of photon intensities like in Fig. \ref{figure_Markov}(b).
 The model, however, breaks down in the case of, e.g., low laser illumination of the chromophores when the photon rates become so low that single photon
  peaks appear in the measurement trace. Binning of photon events into longer time intervals cannot solve the problem as larger time intervals may
   decrease the temporal resolution so strongly that switching can no longer  be observed. 
To solve the problem, we integrate the appearance of single photons
into the Markov model, see Fig. 4. 
State 1 represents the charged QD which does not show laser-induced fluorescence. Consequently, there is no transition possible to state 2, which would represent the charged dot and the presence of a photon in the detector. The random uncharging of the QD at an average rate $\gamma_{\rm out}$ is modeled by a transition to state 3, which represents the uncharged QD. The uncharged QD does exhibit fluorescence and emits photons at an average rate $\gamma_{\rm ph}$ which is modeled by a transition to state 4 that represents the uncharged QD and the presence of a photon in the detector. The photon disappears from the detector at an average detection time $\gamma_{\rm det}$ giving rise to a transition from state 4 back to state 3. While the photon is present in the detector, the QD may change its charging state, giving rise to transitions between state 4 and state 2 with the same rates as in the case of an absent photon (transitions between state 1 and state 3). Finally, the photon disappears from the detector also in the case of a charged quantum dot (state 2) at rate $\gamma_{\rm det}$ which is represented by a transition to state 1.
The model is constructed in such a way that the overall occupation dynamics of the QD is neither influenced by $\gamma_{\rm ph}$  nor by $\gamma_{\rm det}$. The average photon lifetime in the detector is also not influenced by the switching dynamics. We emphasize that the resulting peaks in the measurement trace vary in length as the photon lifetime is exponentially distributed according to $\gamma_{\rm det}$.

\subsection{Quantum polyspectra of Markov dynamics}
\begin{figure}[t]
	\centering
	\includegraphics[width=8.5cm]{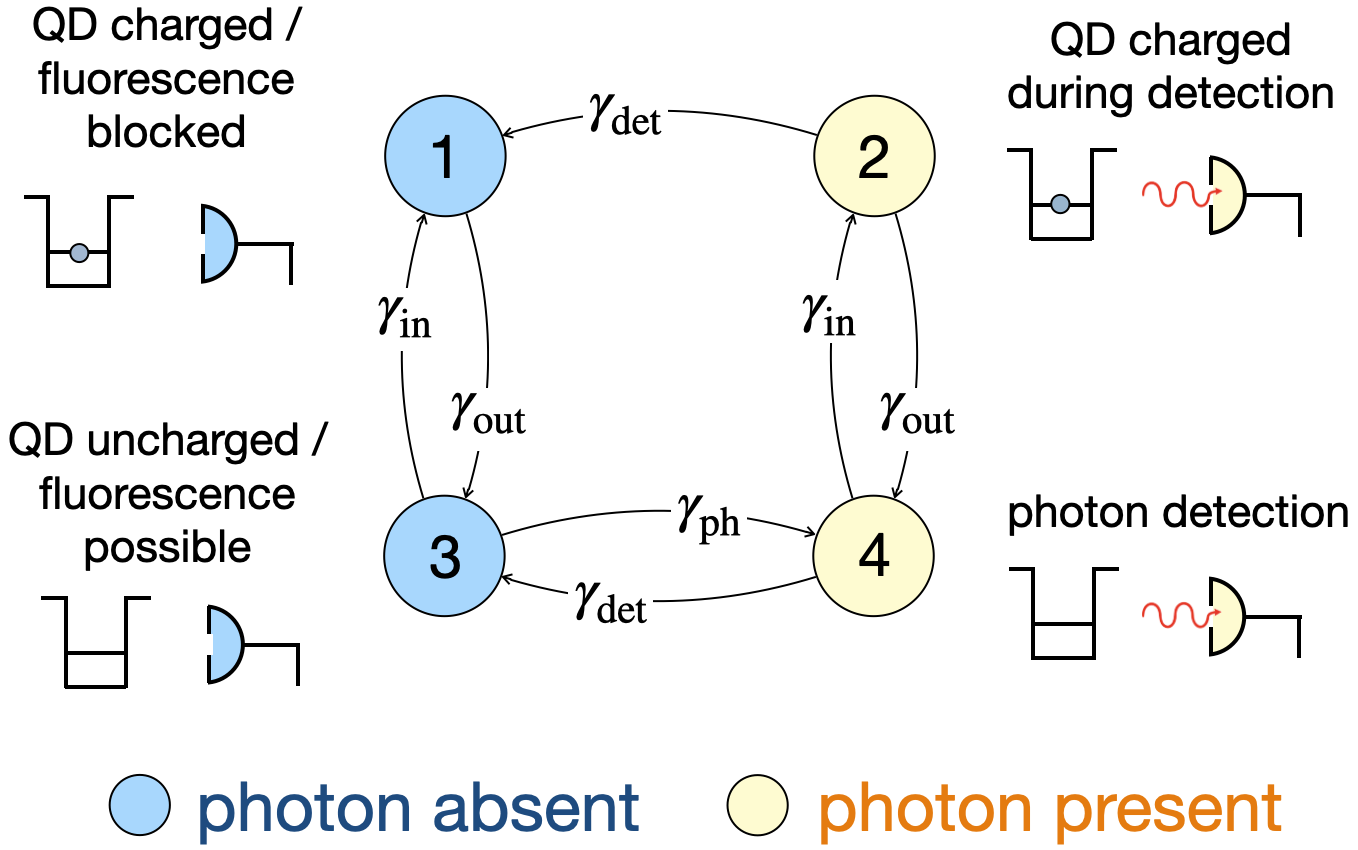}
	\caption{Modified Markov model describing both the stochastic quantum dot occupation and the stochastic emission of a fluorescence photon in the unoccupied state.}
	\label{figure_markov_photon}   
\end{figure} 
In this section we show how Markov dynamics can be treated via a quantum mechanical master equation. Higher order polyspectra of Markov dynamics follow from the powerful framework of continuous quantum measurement theory where recently very general expressions for polyspectra had been found \cite{hagelePRB2018,hagelePRB2020E}.
We represent the four Markov states of Fig. \ref{figure_markov_photon} by quantum states 
$|1 \rangle$, $|2 \rangle$, $|3 \rangle$, and $|4 \rangle$. A Markov system being in a state $| j \rangle$ with probability $p_j$ at time $t$ can then be represented by the density matrix
$\rho(t)  = \sum_j \rho_{jj}(t) |j\rangle \langle j|$, where $\rho_{jj} = p_j$. The dynamics of the Markov system is given by  the transition rates between the states. A transition from, e.g.,  state $|2\rangle$ to state $|4\rangle$  is represented by a jump operator $d = |4\rangle \langle 2|$. The equation of motion for the density matrix is then given by
\begin{equation}
  \dot{\rho} = \gamma {\cal D}[d](\rho),
\end{equation} 
where $\gamma$ is the transition rate and
\begin{equation}
  {\cal D}[d](\rho) = d \rho d^\dagger – (d^\dagger d \rho + \rho d^\dagger d) /2,
\end{equation} 
is a superoperator acting on the density matrix describing the incoherent transition between two states \cite{tilloyPRA2018}.
The dynamics of the full Markov model in Fig. \ref{figure_markov_photon} is reformulated with the help of a Liouvillian ${\cal L}$ acting on the density matrix. A compact formulation for ${\cal L}$ is obtained after introducing the annihilation operator for the electron
\begin{equation}
 a =  |3\rangle \langle 1| + |4\rangle \langle 2|
\end{equation}
and the annihilation operator for the photon
\begin{equation}
 b =  |3\rangle \langle 4| + |1\rangle \langle 2|.
\end{equation}
The equation of motion
\begin{eqnarray}\label{eq:sme2}
	\mathrm{d}\rho &=& \gamma_{\rm in} \mathcal{D}[a^\dagger](\rho) \,dt + 
	\gamma_{\rm out} \mathcal{D}[a](\rho) \,dt \nonumber \\
	&+& \gamma_{\rm ph} \mathcal{D}[(1-a^\dagger a) b^\dagger ](\rho) \,dt + 
	\gamma_{\rm det} \mathcal{D}[b](\rho) \,dt \nonumber \\
		&=& \mathcal{L} \rho \,dt
\end{eqnarray}
covers all of the seven transitions depicted in Fig. \ref{figure_markov_photon}. 
A short calculation yields for the damping operator of the third term  $(1-a^\dagger a) b^\dagger = |4\rangle \langle 3|$, i.e. a transition from state $|3\rangle$ to $|4\rangle$ at rate $\gamma_{\rm ph}$ as required by our model.
Equation (\ref{eq:sme2}) describes how the probabilities $p_j$ of finding the system in state $j$ change over time. 
We emphasize that $\rho(t)$ stays always diagonal, unlike in the general quantum case where coherence between states leads to non-zero 
off-diagonal elements of $\rho(t)$.
The usual master equation does not reproduce the actual stochastic behavior of the system nor can it directly be used to simulate measurement traces $z(t)$  (telegraph noise) like, e.g., the trace shown in Fig. \ref{figure_Markov}(b). The so-called stochastic master equation (SME) is, however, able to describe both the measurement outcome $z(t)$ as well as the stochastic behavior of the system $\rho(t)$ \cite{jacobsCP2006,barchielliNC1982,barchielliBOOK2009,belavkinConf1987, diosiPLA1988,gagenPRA1993,korotkovPRB1999,korotkovPRB2001,goanPRB2001,Attal2006,Attal2010,Gross2018}. Suppose the detector associates  an output voltage $U_j$ to the state $|j \rangle$, then the measurement operator is within the stochastic master equation approach given by 
\begin{equation}
     A = \sum_j U_j |j\rangle \langle j|.
\end{equation}
The measurement output is given by
\begin{equation}
	z(t)  =  \beta^2 {\rm Tr}[\rho(t) (A + A^\dagger)/2]+ \beta \Gamma(t)/2, \label{SME_detector}
\end{equation}
where the output scales with the measurement strength $\beta^2$ and  $\Gamma(t)$ is white background noise, where $\langle \Gamma(t) \Gamma(t') \rangle = \delta(t-t')$. The equation reflects the fact that a weak measurement of $A$ gives information about $A$ that is partly hidden behind background noise. This way a collapse of the quantum state into a definite eigenstate of $A$ is avoided.
In our case of Markov dynamics, the switching behavior (random telegraph noise) is obtained in the strong measurement limit $\beta \gg 1$, where the system reveals its state immediately and the detector output $z(t)$ shows the corresponding voltage level $\beta^2 U_j$ \cite{korotkovPhysRevB2001}.

\begin{figure}[t]
	\centering
	\includegraphics[width=7cm]{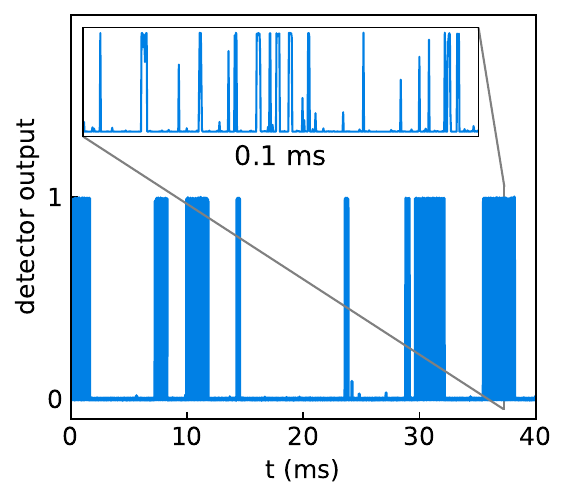}
	\caption{Simulated detector output $z(t)/\beta^2$ obtained from solving the stochastic master equation, Eqs. (\ref{SME_detector}) and (\ref{eq:smeV2}). A clear telegraph like behavior of the dark and bright state is observed. Single photon events appear as peaks of stochastically varying temporal length when zooming into a 1~ms interval of the measurement trace (see inset).
	The stochastic weights of the peaks are regarded via Monte Carlo resampling in the comparison of single click data with model spectra based on the SME (see App. \ref{app:NumericalSpectra}).} 
	\label{fig:simulatedTrace}
\end{figure}

 The system behaves during measurement according to the SME as (notation of \cite{sifftPRA2023})
\begin{eqnarray}
  d \rho & = & {\cal L}\rho \, dt + \beta^2 {\cal D}[A](\rho) \, dt  \nonumber \\
& & + \beta(A \rho + \rho A^\dagger – {\rm Tr}[(A + A^\dagger) \rho ] \rho\,dW, \label{eq:smeV2}
\end{eqnarray}
where the last line describes a stochastic measurement backaction driven by the Wiener-process $W$, where formally $\dot{W} = \Gamma(t)$. Together with the last term in the first line, the equation describes a collapse of the quantum system into an eigenstate of $A+A^\dagger$. The rate of the collapse scales with $\beta^2$. 
According to continuous measurement theory, the spectra of $z(t)$ of a system that is observed in steady state are given in terms of $A$ and ${\cal L}’ = {\cal L}  + \beta^2 {\cal D}[A]$ \cite{hagelePRB2018}.  In the case of pure Markov dynamics, where the density matrix $\rho(t)$ is always diagonal, it is easy to show that the second term in line one disappears and ${\cal L}’ = {\cal L}$. 
Figure \ref{fig:simulatedTrace} shows a simulated measurement trace $z(t)$ of the quantum dot Markov model for $\gamma_{\rm in} = 0.27$~kHz,
$\gamma_{\rm out} = 0.8$~kHz, $\gamma_{\rm ph} = 298$~kHz, $\gamma_{\rm det} = 5000$~kHz, $\beta^2 = 25 000$~kHz, and measurement operator
$A =  |2\rangle \langle 2| +  |4\rangle \langle 4 |$. Traces $z(t)/\beta^2$ were calculated using the QuTiP library \cite{JOHANSSON20131234}. Please note that the single photon events appear as peaks of different lengths and weights. This is a consequence of our Markov model, where the photon can be detected for time intervals of stochastic lengths governed by the photon decay rate $\gamma_{\rm det}$. The trace exhibits several peaks that do not reach the value 1. This is a well-known consequence of the finite measurement strengths $\beta^2$. Only in the ultra-strong measurement limit a trace exhibiting only values of 1 and 0 can be reached (apart from so-called spikes with vanishing weight \cite{TilloyPRA2015}). In contrast, experimental measurement records store the photon arrival times $t_j$ with no information about the peak areas. This discrepancy between experimental records and Markov model is taken care of by a Monte Carlo resampling procedure for calculating polyspectra from single-photon data (see  App. \ref{app:NumericalSpectra}).  

The analytical higher order spectra $S^{(n)}_z$ of $z(t)$ are given in terms of the propagator \cite{hagelePRB2018}
\begin{equation}
{\cal G}(\tau) = e^{{\cal L}'\tau}\Theta(\tau)
\end{equation}  
(with Heaviside-step-function $\Theta(\tau)$),
the steady state
\begin{equation}
  \rho_0 = {\cal G}(\infty) \rho(t),
\end{equation}  
and the measurement superoperator 
\begin{equation}
{\cal A} x = (A x + x A^\dagger)/2.
\end{equation}
The density operator is represented for calculations as an $N \times N$ matrix for $N$ Markov states. The superoperators act linearly on such matrices requiring a representation by $N^4$ numbers.   
Very compact expressions follow after introducing the modified propagator
${\cal G}'(\tau) = {\cal G}(\tau)-  {\cal G}(\infty)\Theta(\tau)$ and the modified measurement operator ${\cal A}' x = {\cal A} x - {\rm Tr}({\cal A}\rho_0)x$. The authors of \cite{hagelePRB2018} found
\begin{eqnarray}
	S_z^{(2)}(\omega) &=& \beta^4 ( {\rm Tr}[{\cal A}'{\cal G}'(\omega){\cal A}'\rho_0]  + {\rm Tr}[{\cal A}'{\cal G}'(-\omega){\cal A}'\rho_0] ) \nonumber \\
	& & \hspace{-0.3cm} + \beta^2/4, 
	\label{eq:S2}
\end{eqnarray}
where ${\cal G}'(\omega) = \int {\cal G}'(\tau) e^{i \omega \tau}\,d\tau$ is the Fourier transform of ${\cal G}'(\tau)$. The third order spectrum is
\begin{eqnarray}		
	S_z^{\rm (3)}(\omega_1,\omega_2,\omega_3 = -\omega_1-\omega_2)  &= & \nonumber \\
	& & \hspace{-40mm}	\beta^6\hspace{-6mm} \sum_{\{k,l,m\} \in \text{prm.} \{1,2,3\}} \hspace{-6mm} {\rm Tr}[{\cal A}'{\cal G}'(\omega_m){\cal A}'{\cal G}'(\omega_m + \omega_l){\cal A}'\rho_0], \label{eq:S3}		
\end{eqnarray}
where the sum regards all six permutations (prm.) of the indices of the $\omega_j$s \cite{footnote1}. The fourth-order spectrum is given in App. \ref{app:QuantumPolyspectra}.
We state for completeness the first order polyspectrum
\begin{eqnarray}
	S_z^{(1)} &=&  \langle z(t) \rangle \nonumber \\
	& = & \beta^2  {\rm Tr}[{\cal A}\rho_0]  	\label{eq:S1},
\end{eqnarray}
which is simply the expectation value of the measurement operator. Usually, $S_z^{(1)}$ is non-zero as can be seen from Figure \ref{fig:simulatedTrace} where all peaks have a positive weight.
\section{Analysis of a quantum emitter}
\label{sec:analysis}
In this section we apply our theory to the analysis of single photon data measured by Kurzmann {\it et al.} on a single InAs quantum dot embedded in a p-i-n diode matrix \cite{kurzmannPRL2019, KurzmannSupp2019}. Random in- and out-tunneling of electrons switches the quantum dot between a charged and uncharged state (see Fig. \ref{figure0}). Our aim is to recover the in- and out-tunneling rates $\gamma_\text{in}$ and $\gamma_\text{out}$ for the case of a strongly reduced fraction $\alpha$ of photons [see Fig. \ref{sample_poly}(a) and Fig. \ref{figure_markov_photon}]. Experimental polyspectra up to fourth order are calculated from the photon arrival times using the recipe of App. \ref{app:NumericalSpectra}. The spectra are displayed in Fig. \ref{sample_poly}(b-d) along with model spectra that were calculated by fitting the corresponding quantum polyspectra
(details see below). The power spectrum $S^{(2)}$ features a peak at zero frequency on top of a flat background. The background arises from the temporally short photon clicks, which in the frequency domain are much broader than the spectral features related to the quantum dot dynamics. The background increases relative to the height of the zero frequency peak as the photon fraction $\alpha$ decreases. Similarly, the bispectrum $S^{(3)}$ and trispectrum $S^{(4)}$ also exhibit an overall flat positive background and some additional structure. This background 
is subtracted in Fig. \ref{sample_poly}(c-d) for a better visualization of the structure. 

The model spectra $S^{(1)}$, $S^{(2)}$, $S^{(3)}$, and $S^{(4)}$  
are fitted simultaneously to their experimental counterparts. The model spectra are obtained from numerically evaluating Eqs. (\ref{eq:S1}), (\ref{eq:S2}), (\ref{eq:S3}), and (\ref{eq:S4}), respectively for the measurement operator ${\cal A}$ and the Liouvillian ${\cal L}$ [Eq. (\ref{eq:sme2})] which depends on the parameters $\gamma_\text{in}$, $\gamma_\text{out}$, $\gamma_\text{ph}$. 
All model spectra were calculated using our QuantumCatch Software library \cite{sifftQuantumCatch2024}.
An analytic evaluation is only possible for very simple Liouvillians ${\cal L}$ which can be diagonalized algebraically. 
 The experimental spectra $S^{(4)}$ and $S^{(3)}$ contribute with $N^2$ data points where $N$ is the number of points used to discretize the spectrum along one axis. The spectrum $S^{(2)}$ contributes with $N$ data points and $S^{(1)}$ with one data point. The weight of each data point that enters the fitting procedure is given by its inverse error (square root of the variance) which is estimated during the calculation of the measured spectra. Data points that appear twice due to symmetry of the spectra are counted only with half their weight. Overall, only the parameters $\gamma_\text{in}$, $\gamma_\text{out}$, and the measurement strength $\beta$ need to be fitted. The photon rate $\gamma_\text{ph}$ is not
an independent fitting parameter. Since $\gamma_\text{in}^{-1}$ is the average time in the bright state of the quantum dot and $\gamma_\text{out}^{-1}$ is the average time in the dark state, the relation $\gamma_\text{ph} \frac{\gamma_\text{in}^{-1}}{\gamma_\text{in}^{-1} + \gamma_\text{out}^{-1}} = N_\text{click}/T_\text{measure}$ allows to estimate $\gamma_\text{ph}$, where $N_\text{click}$ is the number of clicks measured during the overall measurement time  $T_\text{measure}$.
 The detector rate $\gamma_\text{det}$ is fixed to $10^5$~kHz, i.e., it is much faster than the expected photon emission rate. This ensures that the click peak is much shorter than the typical time-interval between the emission of two photons. Moreover, dead times due to the presence of a photon which blocks the emission of a second photon are reduced. 
The area under a single photon peak in the simulated $z(t)$ is the product of
the peak height $\beta^2$ [compare Fig. \ref{fig:simulatedTrace}] and the average temporal length of the peak which is given by the inverse detector transition rate $\gamma_\text{det}^{-1}$. The evaluation scheme of App. \ref{app:NumericalSpectra} assumes unity for the average peak area.  
 The measurement strength $\beta^2$ does in the case of Markov dynamics not enter the Liouvillian ${\cal L}$, but allows for the required scaling  of $z(t)$ and the quantum polyspectra (see Sec. IIIb). The fitting procedure will therefore always yield a $\beta^2 \approx   \gamma_\text{det}$.
The overall structures of the quantum polyspectra do not change for a higher rate $\gamma_\text{det}$ as the spectral features of a single click are unstructured and always much broader than the features of the relevant on-off dynamics of the quantum emitter. 

\begin{figure}[t]
	\centering
	\includegraphics[width=6.6cm]{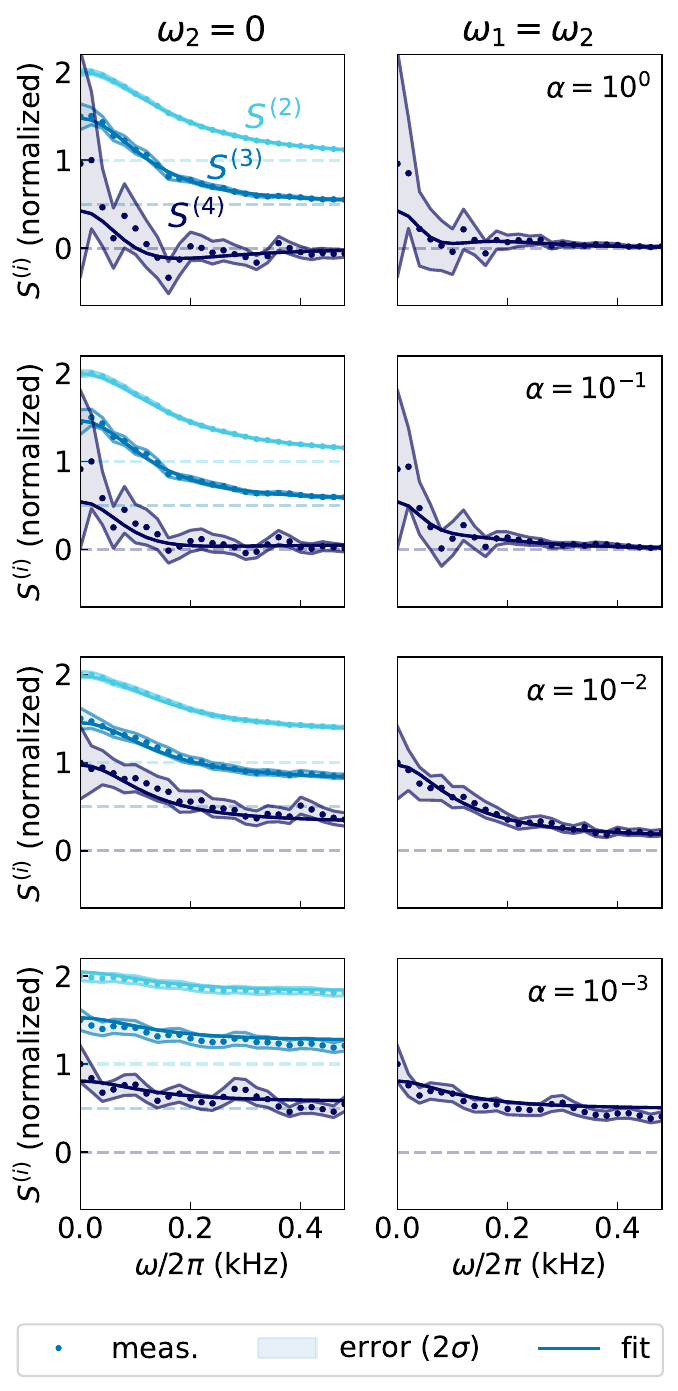}
	\caption{Cuts through experimental polyspectra up to fourth order calculated (dotted line) for photon fractions \mbox{$\alpha = 1,\,10^{-1},\,10^{-2},\,10^{-3}$} in comparison with quantum polyspectra of the model system used for fitting (solid line). The values have been offset for visual clarity, as indicated by the dashed lines.}
	\label{fit}
\end{figure}

\begin{figure}[t]
	\centering
	\includegraphics[width=8cm]{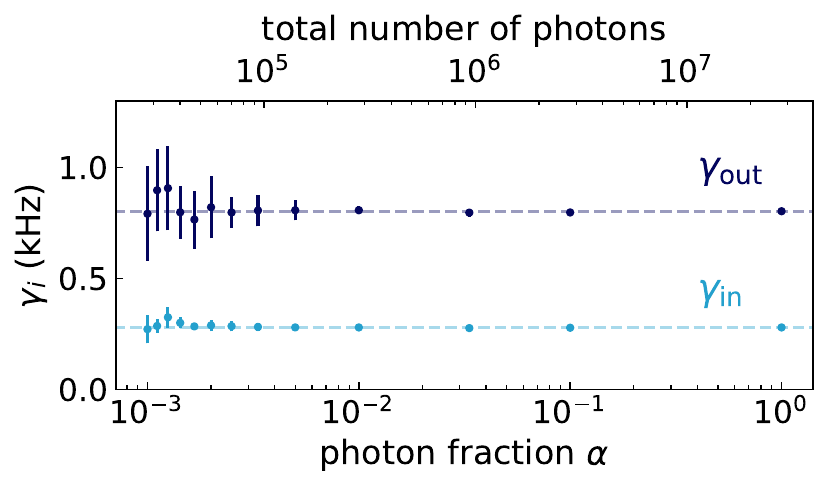}
	\caption{Tunneling rates $\gamma_\text{in}$ and $\gamma_\text{out}$ are obtained from fitting quantum polyspectra to experimental spectra. 
	A photon fraction as low as $\alpha = 10^{-3}$ is still sufficient to fully characterize the blinking behavior. The mean values (dots) and error bars ($\pm 3 \sigma$) are derived from fits to 10 different subsets of the total photon clicks with the same $\alpha$. The estimated tunneling rates for $\alpha = 1$ are shown as dashed lines.}
	\label{fitted_gammas}
\end{figure}

Figure \ref{fit} shows cuts through the experimental spectra $S^{(3)}$ and $S^{(4)}$ along with the power spectrum $S^{(2)}$ for photon fractions between $\alpha = 1$ and $\alpha = 10^{-3}$. The corresponding $\rm 2\sigma$ errors appear as error bands in the plots. All fits of the full spectra are located within the vicinity of the $2 \sigma$ error bands.
This confirms that our Markov model captures correctly all system properties contained in the measured data.
The relative noise on the spectra $S^{(2)}$, $S^{(3)}$, and $S^{(4)}$ strongly increases with the order of the spectrum. Such a behavior is known for all cumulant based quantities \cite{schefczikARXIV2019}. Since $S^{(5)}$ or any spectrum of higher order would exhibit much more noise, it is neither necessary to calculate such spectra nor necessary to regard them in an evaluation procedure. 
Figure \ref{fitted_gammas} shows the $\gamma_\text{in}$ and $\gamma_\text{out}$ tunneling rates that could be determined from the experimental spectra. To assess the reliability of these estimates, the errors of the calculated values were determined by performing our fitting routine on 10 different subsets of the total photon clicks for each photon fraction $\alpha < 1$. The average values of the 10 different pairs of tunneling rates are shown as dots. Their standard deviations $\sigma(\alpha)$ appear as $\pm 3\sigma$ error bars. Useful estimates for the tunneling rates are obtained down to very low photon fractions of $\alpha = 10^{-3}$. 

Therefore, our method can determine transition rates even when photon shot noise dominates the measured spectra. 
The value of the larger tunneling rate also exhibits a larger error. The dependencies of the errors on the system parameters $\gamma_{\rm in}$, $\gamma_{\rm out}$, $\gamma_{\rm ph}$ (which depends on $\alpha$), and the overall measurement time are not trivial as the analytic expressions for $S^{(1)}$ to $S^{(4)}$ which are the fitting functions [see Eqs. (\ref{eq:S2}), (\ref{eq:S3}), (\ref{eq:S1}), (\ref{eq:S4})] depend on the factor $\exp({\cal L}' t)$. This factor depends non-linearly on $\gamma_{\rm in}$, $\gamma_{\rm out}$, $\gamma_{\rm ph}$ which appear as parameters in the Liouvillian ${\cal L}'$ . We nevertheless have started to numerically study the dependencies of the errors. The problem is, however, clearly beyond the scope of the present article.


\section{Discussion}
\label{sec:discussion}
Here, we put our work in context with previous work of quantum measurement theory and with the classical theory of stochastic processes. 
The stochastic emission of photons contains information on the quantum system. We recently gave a theoretical treatment of such “random-time quantum measurements” in terms of quantum polyspectra of general systems that may exhibit both coherent quantum dynamics and damping in Markov approximation \cite{sifftPRA2023}. Their polyspectra were shown to reveal the same information as polyspectra obtained in a traditional continuous quantum measurement. The present article gives a first successful example for analyzing a real-world random-time measurement in the pure Markov case.

In 1955, Cox introduced a class of classical stochastic processes where the rate $\gamma(t)$ of events (like photon clicks) is itself a stochastic process \cite{coxJRSSSB1955}. The blinking dynamics treated here is therefore a special case of a Cox-process, where $\gamma(t)$ originates from a Markov model. Our quantum polyspectra approach is a solution to the question of how the stochastic process $\gamma(t)$ can be fully characterized from data of the original Cox-process. To the best of our knowledge, our approach via polyspectra is unparalleled in the classical theory of Cox-processes. We expect that our approach will inspire more work on the analysis of general Cox-processes. 
We are aware that our current notation is mostly accessible only to researches with a profound background in quantum mechanics as the theory of quantum polyspectra is based on a  sound understanding of quantum mechanical density matrices and their master equations. While this general approach easily allows for the inclusion of quantum dynamics (see \cite{sifftPRA2023}) it asks in the Markov-case for a simplified reformulation in the language of continuous Markov models. Since $\rho$ is always diagonal it can be represented by a vector with only $N$ entries. The Liouvillian ${\cal L}$ (represented by $N^4$ numbers) would be replaced by a much simpler and numerically less demanding $N \times N$ transition matrix.    
Also, the equations for analytical polyspectra would no longer relate to quantum objects but rather to entities of Markov theory making our polyspectra approach more accessible to researches from mathematical statistics. 

\section{Conclusion} 

In this study, we have introduced an evaluation scheme that leverages higher-order spectra of single-photon measurements to analyze the blinking dynamics of quantum emitters. Our method eliminates the notorious requirement for a minimum light level, as the spectra can be calculated directly from the single-photon measurements. This allows us to extract valuable information about blinking dynamics in regimes where traditional methods based on photon binning would fail. Our approach is highly versatile and can be readily generalized to multi-state Markov models and quantum systems exhibiting coherent dynamics. This opens the door to exploring even more complex quantum phenomena, such as the transition to the quantum Zeno regime under single-photon measurements \cite{sifftPRA2023}.

Generally, our scheme does not compromise measurement bandwidth or the accuracy of recovered system parameters, since the system dynamic frequencies that can be recovered from single-photon measurements are not constrained by the average photon rate. This enables the investigation of systems previously considered beyond analytical reach and could inspire the design of experiments deliberately in the low photon rate regime, where disturbances to quantum systems can be minimized. This innovation paves the way for new research opportunities in nano-electronics, quantum sensing, and fluorescence microscopy.

\begin{acknowledgments}
	We acknowledge financial support by the German Science Foundation (DFG) under Project Nos. 341960391 and 510607185 (D.H.),
	 383065199 (A. Lu. and M.G.), 278162697 (A. Lo. and M. G. within SFB 1242) as well as by the Mercator Research Center Ruhr under Project No. Ko-2022-0013.
\end{acknowledgments}

\appendix
\section{Polyspectra of single photon measurements via Monte Carlo resampling}
\label{app:NumericalSpectra}
The calculation of polyspectra from bandwidth-limited data is based on the sampled data $z_j$, see Ref. \cite{sifftPRR2021}. The data stream $z(t)$ is divided into time-intervals of length $T$ corresponding 
to $N$ data points $z_j^{(n)}$,  where $n$ is the number of the interval and $j = 0,…,N-1$. The Fourier-coefficients
\begin{equation}
   a_k^{(n)} = \frac{T}{N} \sum_{j=0}^{N-1} g_j  z_j^{(n)}   e^{2 \pi i j k / N}
\end{equation}
are the basis for estimating the power spectrum
\begin{equation}
  S_z^{(2)}(\omega_k = 2 \pi k /T) \approx \frac{N C_2(a^{(n)}_k,(a^{(n)}_k)^*)}{T \sum_j g_j^* g_j}, \label{eq:AppAspec}
\end{equation}
where $g_j$ are the coefficients of a so-called window function which improves the spectral resolution of the spectrum \cite{sifftPRR2021,starosielecSP2014}, and  
$C_2(x,y) = \langle xy \rangle - \langle x \rangle \langle y \rangle$ is the second order cumulant (identical with the covariance). 
The expectation values $\langle … \rangle$ refer to the ideal case of an infinite amount of data. 
In case of a limited number $m$ of data pairs $x$ and $y$, 
the estimator $c_2(x,y) = \frac{m}{m-1}(\overline{xy}- \overline{x}\,\overline{y})$ yields an unbiased estimate of $C_2(x,y)$ where the overline  $\overline{(…)}$ denotes the average of $m$ samples \cite{schefczikARXIV2019}.  
The famous prefactor $1/(m-1)$ is known as the Bessel-correction and appears in literature for estimators of the variance (see Ref. \cite{fisherPLMS1930} for the variance estimator and its higher order generalizations).
Corresponding expression for the bispectrum $S_z^{(3)}$ and trispectrum $S_z^{(4)}$ can be found in Appendix B of \cite{sifftPRR2021}. 
The analysis of single-photon click-events requires a modification of the scheme above for two reasons: (i) Photons detected at times $t_j$ correspond to a continuous measurement record $z(t) = \sum_j \delta(t -t_j)$ where $\delta(\tau)$ is the Dirac-delta distribution-function. A naive discretization of $z(t)$ into finite time-steps is no longer possible. (ii) The detector model within our Markov-theory does not yield delta-pulses for the theoretical detector output $z(t)$. Instead, short box-shaped pulses of varying temporal length $\Delta t$ appear, where $\Delta t$ is exponentially distributed with
the distribution function $p(\Delta t) = \gamma_{\rm det} e^{-\gamma_{\rm det}\Delta t}$ for $\Delta t >0$ and $0$ otherwise. The contribution of such pulses to the Fourier coefficient therefore varies correspondingly. 
For a comparison of experimental spectra and model spectra the issues (i) and (ii) have to be regarded. Consider click times $t^{(n)}_j$ that relate to the same time interval $n$ of length $T$. A Monte Carlo resampling of the click events yield new Fourier coefficients
\begin{equation}
  {a’}_k^{(n)} = \sum_j g(t^{(n)}_j) b^{(n)}_j \exp(i \omega_k t_j^{(n)}),
\end{equation}
where the $\delta$-like contribution at time $t_j^{(n)}$ enter the coefficient and the exponential distribution is regarded via the new random variables $b^{(n)}_j$ which are distributed according to $p(x) = e^{-x}$. The polyspectra are then calculated via Eq. (\ref{eq:AppAspec}) and its higher order generalizations by replacing $a_k^{(n)}$ by ${a’}_k^{(n)}$ and averaging over typically 100 different realizations of random $b^{(n)}_j$s. 
A naive approach, where $b^{(n)}_j \equiv 1$, would result in incorrect spectra as higher order moments of $b_j$ appear in the calculation of spectra and the correct exponential distribution yields, e.g.,  
$\langle b^{(n)}_j  \rangle = 1$, $\langle (b^{(n)}_j)^2  \rangle = 2$,   $\langle (b^{(n)}_j)^3  \rangle = 3!$ while $b^{(n)}_j \equiv 1$ would yield the value 1 for all those moments.
All experimental spectra have been calculated using our SignalSnap toolbox \cite{sifftSIGNALSNAP2022}.
While the introduction of new randomness via Monte Carlo resampling seems not to be very elegant, its numerical implementation is straight forward and bridges successfully experiment and theory. Nevertheless, a more direct scheme of calculating experimental spectra from the time stamps $t_j$ 
without the need for introducing new randomness is certainly desirable but currently elusive.


\section{Fourth-order quantum polyspectrum}
\label{app:QuantumPolyspectra}
The fourth-order polyspectrum of the detector output $z(t)$  of the continuously monitored quantum system in the steady state follows from the SME
and the definition of  Brillinger's polyspectra $S_z^{(n)}$. 
The spectrum (second- and third-order spectra see main text) \cite{footnote1}
\begin{widetext}
\begin{eqnarray}	
	S_z^{\rm (4)}(\omega_1,\omega_2,\omega_3,\omega_4 = -\omega_1-\omega_2-\omega_3) & = & \hspace{4mm}\beta^8 \hspace{-8mm} \sum_{\{k,l,m,n\} \in \text{prm.} \{1,2,3,4\}}
	\hspace{-8mm} \left[ {\rm Tr}[{\cal A}'{\cal G}'(\omega_n){\cal A}' {\cal G}'(\omega_m + \omega_n){\cal A}'{\cal G}'(\omega_l + \omega_m + \omega_n){\cal A}'\rho_0] \right.  \label{eq:S4} \\  \nonumber
	&-& \frac{1}{2 \pi}\int{\rm Tr}[{\cal A}'{\cal G}'(\omega_n) {\cal G}'(\omega_m + \omega_n - \omega){\cal A}'\rho_0]{\rm Tr}[{\cal A}'{\cal G}'(\omega) {\cal G}'(\omega_l +\omega_m + \omega_n){\cal A}'\rho_0]\textrm{d}\omega \\ \nonumber
	&-& \left. \frac{1}{2 \pi}\int{\rm Tr}[{\cal A}'{\cal G}'(\omega_n) {\cal G}'(\omega_l +\omega_m + \omega_n){\cal G}'(\omega_m + \omega_n - \omega){\cal A}'\rho_0]{\rm Tr}[{\cal A}'{\cal G}'(\omega) {\cal A}'\rho_0]\textrm{d}\omega\right]. \label{eq:S4}
\end{eqnarray}
was first derived in Refs. \cite{hagelePRB2018,hagelePRB2020E} where also an efficient methods for its evaluation was given.  
The spectra of this work were numerically calculated from the Liouvillian ${\cal L}$ and measurement operator ${\cal A}$ using our software library QuantumCatch which is based on the QuTiP and ArrayFire software libraries \cite{sifftQuantumCatch2024,JOHANSSON20131234,Yalamanchili2015}
\end{widetext}

\bibliography{spinnoiseV2}

\begin{thebibliography}{56}%
\makeatletter
\providecommand \@ifxundefined [1]{%
 \@ifx{#1\undefined}
}%
\providecommand \@ifnum [1]{%
 \ifnum #1\expandafter \@firstoftwo
 \else \expandafter \@secondoftwo
 \fi
}%
\providecommand \@ifx [1]{%
 \ifx #1\expandafter \@firstoftwo
 \else \expandafter \@secondoftwo
 \fi
}%
\providecommand \natexlab [1]{#1}%
\providecommand \enquote  [1]{``#1''}%
\providecommand \bibnamefont  [1]{#1}%
\providecommand \bibfnamefont [1]{#1}%
\providecommand \citenamefont [1]{#1}%
\providecommand \href@noop [0]{\@secondoftwo}%
\providecommand \href [0]{\begingroup \@sanitize@url \@href}%
\providecommand \@href[1]{\@@startlink{#1}\@@href}%
\providecommand \@@href[1]{\endgroup#1\@@endlink}%
\providecommand \@sanitize@url [0]{\catcode `\\12\catcode `\$12\catcode
  `\&12\catcode `\#12\catcode `\^12\catcode `\_12\catcode `\%12\relax}%
\providecommand \@@startlink[1]{}%
\providecommand \@@endlink[0]{}%
\providecommand \url  [0]{\begingroup\@sanitize@url \@url }%
\providecommand \@url [1]{\endgroup\@href {#1}{\urlprefix }}%
\providecommand \urlprefix  [0]{URL }%
\providecommand \Eprint [0]{\href }%
\providecommand \doibase [0]{https://doi.org/}%
\providecommand \selectlanguage [0]{\@gobble}%
\providecommand \bibinfo  [0]{\@secondoftwo}%
\providecommand \bibfield  [0]{\@secondoftwo}%
\providecommand \translation [1]{[#1]}%
\providecommand \BibitemOpen [0]{}%
\providecommand \bibitemStop [0]{}%
\providecommand \bibitemNoStop [0]{.\EOS\space}%
\providecommand \EOS [0]{\spacefactor3000\relax}%
\providecommand \BibitemShut  [1]{\csname bibitem#1\endcsname}%
\let\auto@bib@innerbib\@empty
\bibitem [{\citenamefont {Stevens}(2013)}]{stevensREVIEW2013}%
  \BibitemOpen
  \bibfield  {author} {\bibinfo {author} {\bibfnamefont {M.~J.}\ \bibnamefont
  {Stevens}},\ }\bibfield  {title} {\bibinfo {title} {Photon statistics,
  measurements, and measurements tools},\ }\href@noop {} {\bibfield  {journal}
  {\bibinfo  {journal} {Experimental Methods in the Physical Sciences}\
  }\textbf {\bibinfo {volume} {45}},\ \bibinfo {pages} {25} (\bibinfo {year}
  {2013})}\BibitemShut {NoStop}%
\bibitem [{\citenamefont {Andrec}\ \emph {et~al.}(2003)\citenamefont {Andrec},
  \citenamefont {Levy},\ and\ \citenamefont {Talaga}}]{AndrecPCA2003}%
  \BibitemOpen
  \bibfield  {author} {\bibinfo {author} {\bibfnamefont {M.}~\bibnamefont
  {Andrec}}, \bibinfo {author} {\bibfnamefont {R.~M.}\ \bibnamefont {Levy}},\
  and\ \bibinfo {author} {\bibfnamefont {D.~S.}\ \bibnamefont {Talaga}},\
  }\bibfield  {title} {\bibinfo {title} {Direct determination of kinetic rates
  from single-molecule photon arrival trajectories using hidden {Markov}
  models},\ }\href {https://doi.org/10.1021/jp035514+} {\bibfield  {journal}
  {\bibinfo  {journal} {The Journal of Physical Chemistry A}\ }\textbf
  {\bibinfo {volume} {107}},\ \bibinfo {pages} {7454} (\bibinfo {year}
  {2003})}\BibitemShut {NoStop}%
\bibitem [{\citenamefont {McKinney}\ \emph {et~al.}(2006)\citenamefont
  {McKinney}, \citenamefont {Joo},\ and\ \citenamefont
  {Ha}}]{mckinneyBIOPHYSJ2006}%
  \BibitemOpen
  \bibfield  {author} {\bibinfo {author} {\bibfnamefont {S.~A.}\ \bibnamefont
  {McKinney}}, \bibinfo {author} {\bibfnamefont {C.}~\bibnamefont {Joo}},\ and\
  \bibinfo {author} {\bibfnamefont {T.}~\bibnamefont {Ha}},\ }\bibfield
  {title} {\bibinfo {title} {Analysis of single-molecule {FRET} trajectories
  using {Hidden Markov Modeling}},\ }\href
  {https://doi.org/10.1529/biophysj.106.082487} {\bibfield  {journal} {\bibinfo
   {journal} {Biophys. J.}\ }\textbf {\bibinfo {volume} {91}},\ \bibinfo
  {pages} {1941} (\bibinfo {year} {2006})}\BibitemShut {NoStop}%
\bibitem [{\citenamefont {J\"ager}\ \emph {et~al.}(2009)\citenamefont
  {J\"ager}, \citenamefont {Kiel}, \citenamefont {Herten},\ and\ \citenamefont
  {Hamprecht}}]{JagerChemPhysChem2009}%
  \BibitemOpen
  \bibfield  {author} {\bibinfo {author} {\bibfnamefont {M.}~\bibnamefont
  {J\"ager}}, \bibinfo {author} {\bibfnamefont {A.}~\bibnamefont {Kiel}},
  \bibinfo {author} {\bibfnamefont {D.-P.}\ \bibnamefont {Herten}},\ and\
  \bibinfo {author} {\bibfnamefont {F.~A.}\ \bibnamefont {Hamprecht}},\
  }\bibfield  {title} {\bibinfo {title} {Analysis of single-molecule
  fluorescence spectroscopic data with a {Markov}-modulated {Poisson}
  process},\ }\href {https://doi.org/https://doi.org/10.1002/cphc.200900331}
  {\bibfield  {journal} {\bibinfo  {journal} {ChemPhysChem}\ }\textbf {\bibinfo
  {volume} {10}},\ \bibinfo {pages} {2486} (\bibinfo {year}
  {2009})}\BibitemShut {NoStop}%
\bibitem [{\citenamefont {G{\"o}tz}\ \emph {et~al.}(2022)\citenamefont
  {G{\"o}tz}, \citenamefont {Barth}, \citenamefont {Bohr}, \citenamefont
  {B{\"o}rner}, \citenamefont {Chen}, \citenamefont {Cordes}, \citenamefont
  {Erie}, \citenamefont {Gebhardt}, \citenamefont {Hadzic}, \citenamefont
  {Hamilton} \emph {et~al.}}]{gotz2022blind}%
  \BibitemOpen
  \bibfield  {author} {\bibinfo {author} {\bibfnamefont {M.}~\bibnamefont
  {G{\"o}tz}}, \bibinfo {author} {\bibfnamefont {A.}~\bibnamefont {Barth}},
  \bibinfo {author} {\bibfnamefont {S.~S.-R.}\ \bibnamefont {Bohr}}, \bibinfo
  {author} {\bibfnamefont {R.}~\bibnamefont {B{\"o}rner}}, \bibinfo {author}
  {\bibfnamefont {J.}~\bibnamefont {Chen}}, \bibinfo {author} {\bibfnamefont
  {T.}~\bibnamefont {Cordes}}, \bibinfo {author} {\bibfnamefont {D.~A.}\
  \bibnamefont {Erie}}, \bibinfo {author} {\bibfnamefont {C.}~\bibnamefont
  {Gebhardt}}, \bibinfo {author} {\bibfnamefont {M.~C.}\ \bibnamefont
  {Hadzic}}, \bibinfo {author} {\bibfnamefont {G.~L.}\ \bibnamefont
  {Hamilton}}, \emph {et~al.},\ }\bibfield  {title} {\bibinfo {title} {A blind
  benchmark of analysis tools to infer kinetic rate constants from
  single-molecule {FRET} trajectories},\ }\href@noop {} {\bibfield  {journal}
  {\bibinfo  {journal} {Nature Commun.}\ }\textbf {\bibinfo {volume} {13}},\
  \bibinfo {pages} {5402} (\bibinfo {year} {2022})}\BibitemShut {NoStop}%
\bibitem [{\citenamefont {Senellart}\ \emph {et~al.}(2017)\citenamefont
  {Senellart}, \citenamefont {Solomon},\ and\ \citenamefont
  {White}}]{senellartNATURE2017}%
  \BibitemOpen
  \bibfield  {author} {\bibinfo {author} {\bibfnamefont {P.}~\bibnamefont
  {Senellart}}, \bibinfo {author} {\bibfnamefont {G.}~\bibnamefont {Solomon}},\
  and\ \bibinfo {author} {\bibfnamefont {A.}~\bibnamefont {White}},\ }\bibfield
   {title} {\bibinfo {title} {High-performance semiconductor quantum-dot
  single-photon sources},\ }\href@noop {} {\bibfield  {journal} {\bibinfo
  {journal} {Nature Nanotech.}\ }\textbf {\bibinfo {volume} {12}},\ \bibinfo
  {pages} {1026} (\bibinfo {year} {2017})}\BibitemShut {NoStop}%
\bibitem [{\citenamefont {Gschrey}\ \emph {et~al.}(2015)\citenamefont
  {Gschrey}, \citenamefont {Thoma}, \citenamefont {Schnauber}, \citenamefont
  {Seifried}, \citenamefont {Schmidt}, \citenamefont {Wohlfeil}, \citenamefont
  {Kr{\"u}ger}, \citenamefont {Schulze}, \citenamefont {Heindel}, \citenamefont
  {Burger}, \citenamefont {Schmidt}, \citenamefont {Strittmatter},
  \citenamefont {Rodt},\ and\ \citenamefont
  {Reitzenstein}}]{GschreyNatComm2015}%
  \BibitemOpen
  \bibfield  {author} {\bibinfo {author} {\bibfnamefont {M.}~\bibnamefont
  {Gschrey}}, \bibinfo {author} {\bibfnamefont {A.}~\bibnamefont {Thoma}},
  \bibinfo {author} {\bibfnamefont {P.}~\bibnamefont {Schnauber}}, \bibinfo
  {author} {\bibfnamefont {M.}~\bibnamefont {Seifried}}, \bibinfo {author}
  {\bibfnamefont {R.}~\bibnamefont {Schmidt}}, \bibinfo {author} {\bibfnamefont
  {B.}~\bibnamefont {Wohlfeil}}, \bibinfo {author} {\bibfnamefont
  {L.}~\bibnamefont {Kr{\"u}ger}}, \bibinfo {author} {\bibfnamefont {J.~H.}\
  \bibnamefont {Schulze}}, \bibinfo {author} {\bibfnamefont {T.}~\bibnamefont
  {Heindel}}, \bibinfo {author} {\bibfnamefont {S.}~\bibnamefont {Burger}},
  \bibinfo {author} {\bibfnamefont {F.}~\bibnamefont {Schmidt}}, \bibinfo
  {author} {\bibfnamefont {A.}~\bibnamefont {Strittmatter}}, \bibinfo {author}
  {\bibfnamefont {S.}~\bibnamefont {Rodt}},\ and\ \bibinfo {author}
  {\bibfnamefont {S.}~\bibnamefont {Reitzenstein}},\ }\bibfield  {title}
  {\bibinfo {title} {Highly indistinguishable photons from deterministic
  quantum-dot microlenses utilizing three-dimensional in situ electron-beam
  lithography},\ }\href {https://doi.org/10.1038/ncomms8662} {\bibfield
  {journal} {\bibinfo  {journal} {Nature Communications}\ }\textbf {\bibinfo
  {volume} {6}},\ \bibinfo {pages} {7662} (\bibinfo {year} {2015})}\BibitemShut
  {NoStop}%
\bibitem [{\citenamefont {Heindel}\ \emph {et~al.}(2012)\citenamefont
  {Heindel}, \citenamefont {Kessler}, \citenamefont {Rau}, \citenamefont
  {Schneider}, \citenamefont {Fürst}, \citenamefont {Hargart}, \citenamefont
  {Schulz}, \citenamefont {Eichfelder}, \citenamefont {Roßbach}, \citenamefont
  {Nauerth}, \citenamefont {Lermer}, \citenamefont {Weier}, \citenamefont
  {Jetter}, \citenamefont {Kamp}, \citenamefont {Reitzenstein}, \citenamefont
  {Höfling}, \citenamefont {Michler}, \citenamefont {Weinfurter},\ and\
  \citenamefont {Forchel}}]{HeindelNJP2012}%
  \BibitemOpen
  \bibfield  {author} {\bibinfo {author} {\bibfnamefont {T.}~\bibnamefont
  {Heindel}}, \bibinfo {author} {\bibfnamefont {C.~A.}\ \bibnamefont
  {Kessler}}, \bibinfo {author} {\bibfnamefont {M.}~\bibnamefont {Rau}},
  \bibinfo {author} {\bibfnamefont {C.}~\bibnamefont {Schneider}}, \bibinfo
  {author} {\bibfnamefont {M.}~\bibnamefont {Fürst}}, \bibinfo {author}
  {\bibfnamefont {F.}~\bibnamefont {Hargart}}, \bibinfo {author} {\bibfnamefont
  {W.-M.}\ \bibnamefont {Schulz}}, \bibinfo {author} {\bibfnamefont
  {M.}~\bibnamefont {Eichfelder}}, \bibinfo {author} {\bibfnamefont
  {R.}~\bibnamefont {Roßbach}}, \bibinfo {author} {\bibfnamefont
  {S.}~\bibnamefont {Nauerth}}, \bibinfo {author} {\bibfnamefont
  {M.}~\bibnamefont {Lermer}}, \bibinfo {author} {\bibfnamefont
  {H.}~\bibnamefont {Weier}}, \bibinfo {author} {\bibfnamefont
  {M.}~\bibnamefont {Jetter}}, \bibinfo {author} {\bibfnamefont
  {M.}~\bibnamefont {Kamp}}, \bibinfo {author} {\bibfnamefont {S.}~\bibnamefont
  {Reitzenstein}}, \bibinfo {author} {\bibfnamefont {S.}~\bibnamefont
  {Höfling}}, \bibinfo {author} {\bibfnamefont {P.}~\bibnamefont {Michler}},
  \bibinfo {author} {\bibfnamefont {H.}~\bibnamefont {Weinfurter}},\ and\
  \bibinfo {author} {\bibfnamefont {A.}~\bibnamefont {Forchel}},\ }\bibfield
  {title} {\bibinfo {title} {Quantum key distribution using quantum dot
  single-photon emitting diodes in the red and near infrared spectral range},\
  }\href {https://doi.org/10.1088/1367-2630/14/8/083001} {\bibfield  {journal}
  {\bibinfo  {journal} {New Journal of Physics}\ }\textbf {\bibinfo {volume}
  {14}},\ \bibinfo {pages} {083001} (\bibinfo {year} {2012})}\BibitemShut
  {NoStop}%
\bibitem [{\citenamefont {M{\"u}ller}\ \emph {et~al.}(2014)\citenamefont
  {M{\"u}ller}, \citenamefont {Bounouar}, \citenamefont {J{\"o}ns},
  \citenamefont {Gl{\"a}ssl},\ and\ \citenamefont
  {Michler}}]{MuellerNatPho2014}%
  \BibitemOpen
  \bibfield  {author} {\bibinfo {author} {\bibfnamefont {M.}~\bibnamefont
  {M{\"u}ller}}, \bibinfo {author} {\bibfnamefont {S.}~\bibnamefont
  {Bounouar}}, \bibinfo {author} {\bibfnamefont {K.~D.}\ \bibnamefont
  {J{\"o}ns}}, \bibinfo {author} {\bibfnamefont {M.}~\bibnamefont
  {Gl{\"a}ssl}},\ and\ \bibinfo {author} {\bibfnamefont {P.}~\bibnamefont
  {Michler}},\ }\bibfield  {title} {\bibinfo {title} {On-demand generation of
  indistinguishable polarization-entangled photon pairs},\ }\href
  {https://doi.org/10.1038/nphoton.2013.377} {\bibfield  {journal} {\bibinfo
  {journal} {Nature Photonics}\ }\textbf {\bibinfo {volume} {8}},\ \bibinfo
  {pages} {224} (\bibinfo {year} {2014})}\BibitemShut {NoStop}%
\bibitem [{\citenamefont {Kurzmann}\ \emph {et~al.}(2019)\citenamefont
  {Kurzmann}, \citenamefont {Stegmann}, \citenamefont {Kerski}, \citenamefont
  {Ludwig}, \citenamefont {Wieck}, \citenamefont {K\"onig}, \citenamefont
  {Lorke},\ and\ \citenamefont {Geller}}]{kurzmannPRL2019}%
  \BibitemOpen
  \bibfield  {author} {\bibinfo {author} {\bibfnamefont {A.}~\bibnamefont
  {Kurzmann}}, \bibinfo {author} {\bibfnamefont {P.}~\bibnamefont {Stegmann}},
  \bibinfo {author} {\bibfnamefont {J.}~\bibnamefont {Kerski}}, \bibinfo
  {author} {\bibfnamefont {A.}~\bibnamefont {Ludwig}}, \bibinfo {author}
  {\bibfnamefont {A.~D.}\ \bibnamefont {Wieck}}, \bibinfo {author}
  {\bibfnamefont {J.}~\bibnamefont {K\"onig}}, \bibinfo {author} {\bibfnamefont
  {A.}~\bibnamefont {Lorke}},\ and\ \bibinfo {author} {\bibfnamefont
  {M.}~\bibnamefont {Geller}},\ }\bibfield  {title} {\bibinfo {title} {Optical
  detection of single-electron tunneling into a semiconductor quantum dot},\
  }\href {https://doi.org/10.1103/PhysRevLett.122.247403} {\bibfield  {journal}
  {\bibinfo  {journal} {Phys. Rev. Lett.}\ }\textbf {\bibinfo {volume} {122}},\
  \bibinfo {pages} {247403} (\bibinfo {year} {2019})}\BibitemShut {NoStop}%
\bibitem [{\citenamefont {Vamivakas}\ \emph {et~al.}(2010)\citenamefont
  {Vamivakas}, \citenamefont {Lu}, \citenamefont {Matthiesen}, \citenamefont
  {Zhao}, \citenamefont {F{\"a}lt}, \citenamefont {Badolato},\ and\
  \citenamefont {Atat{\"u}re}}]{VamivakasNATURE2010}%
  \BibitemOpen
  \bibfield  {author} {\bibinfo {author} {\bibfnamefont {A.~N.}\ \bibnamefont
  {Vamivakas}}, \bibinfo {author} {\bibfnamefont {C.~Y.}\ \bibnamefont {Lu}},
  \bibinfo {author} {\bibfnamefont {C.}~\bibnamefont {Matthiesen}}, \bibinfo
  {author} {\bibfnamefont {Y.}~\bibnamefont {Zhao}}, \bibinfo {author}
  {\bibfnamefont {S.}~\bibnamefont {F{\"a}lt}}, \bibinfo {author}
  {\bibfnamefont {A.}~\bibnamefont {Badolato}},\ and\ \bibinfo {author}
  {\bibfnamefont {M.}~\bibnamefont {Atat{\"u}re}},\ }\bibfield  {title}
  {\bibinfo {title} {Observation of spin-dependent quantum jumps via quantum
  dot resonance fluorescence},\ }\href {https://doi.org/10.1038/nature09359}
  {\bibfield  {journal} {\bibinfo  {journal} {Nature}\ }\textbf {\bibinfo
  {volume} {467}},\ \bibinfo {pages} {297} (\bibinfo {year}
  {2010})}\BibitemShut {NoStop}%
\bibitem [{\citenamefont {Kuhlmann}\ \emph {et~al.}(2013)\citenamefont
  {Kuhlmann}, \citenamefont {Houel}, \citenamefont {Ludwig}, \citenamefont
  {Greuter}, \citenamefont {Reuter}, \citenamefont {Wieck}, \citenamefont
  {Poggio},\ and\ \citenamefont {Warburton}}]{KuhlmannNATPHY2013}%
  \BibitemOpen
  \bibfield  {author} {\bibinfo {author} {\bibfnamefont {A.~V.}\ \bibnamefont
  {Kuhlmann}}, \bibinfo {author} {\bibfnamefont {J.}~\bibnamefont {Houel}},
  \bibinfo {author} {\bibfnamefont {A.}~\bibnamefont {Ludwig}}, \bibinfo
  {author} {\bibfnamefont {L.}~\bibnamefont {Greuter}}, \bibinfo {author}
  {\bibfnamefont {D.}~\bibnamefont {Reuter}}, \bibinfo {author} {\bibfnamefont
  {A.~D.}\ \bibnamefont {Wieck}}, \bibinfo {author} {\bibfnamefont
  {M.}~\bibnamefont {Poggio}},\ and\ \bibinfo {author} {\bibfnamefont {R.~J.}\
  \bibnamefont {Warburton}},\ }\bibfield  {title} {\bibinfo {title} {Charge
  noise and spin noise in a semiconductor quantum device},\ }\href
  {https://doi.org/10.1038/nphys2688} {\bibfield  {journal} {\bibinfo
  {journal} {Nature Physics}\ }\textbf {\bibinfo {volume} {9}},\ \bibinfo
  {pages} {570} (\bibinfo {year} {2013})}\BibitemShut {NoStop}%
\bibitem [{\citenamefont {Levitov}\ \emph {et~al.}(1996)\citenamefont
  {Levitov}, \citenamefont {Lee},\ and\ \citenamefont
  {Lesovik}}]{levitovJMP1996}%
  \BibitemOpen
  \bibfield  {author} {\bibinfo {author} {\bibfnamefont {L.~S.}\ \bibnamefont
  {Levitov}}, \bibinfo {author} {\bibfnamefont {H.}~\bibnamefont {Lee}},\ and\
  \bibinfo {author} {\bibfnamefont {G.~B.}\ \bibnamefont {Lesovik}},\
  }\bibfield  {title} {\bibinfo {title} {Electron counting statistics and
  coherent states of electric current},\ }\href
  {https://doi.org/10.1063/1.531672} {\bibfield  {journal} {\bibinfo  {journal}
  {J. Math. Phys.}\ }\textbf {\bibinfo {volume} {37}},\ \bibinfo {pages} {4845}
  (\bibinfo {year} {1996})}\BibitemShut {NoStop}%
\bibitem [{\citenamefont {Sifft}\ \emph {et~al.}(2021)\citenamefont {Sifft},
  \citenamefont {Kurzmann}, \citenamefont {Kerski}, \citenamefont {Schott},
  \citenamefont {Ludwig}, \citenamefont {Wieck}, \citenamefont {Lorke},
  \citenamefont {Geller},\ and\ \citenamefont {H\"agele}}]{sifftPRR2021}%
  \BibitemOpen
  \bibfield  {author} {\bibinfo {author} {\bibfnamefont {M.}~\bibnamefont
  {Sifft}}, \bibinfo {author} {\bibfnamefont {A.}~\bibnamefont {Kurzmann}},
  \bibinfo {author} {\bibfnamefont {J.}~\bibnamefont {Kerski}}, \bibinfo
  {author} {\bibfnamefont {R.}~\bibnamefont {Schott}}, \bibinfo {author}
  {\bibfnamefont {A.}~\bibnamefont {Ludwig}}, \bibinfo {author} {\bibfnamefont
  {A.~D.}\ \bibnamefont {Wieck}}, \bibinfo {author} {\bibfnamefont
  {A.}~\bibnamefont {Lorke}}, \bibinfo {author} {\bibfnamefont
  {M.}~\bibnamefont {Geller}},\ and\ \bibinfo {author} {\bibfnamefont
  {D.}~\bibnamefont {H\"agele}},\ }\bibfield  {title} {\bibinfo {title}
  {Quantum polyspectra for modeling and evaluating quantum transport
  measurements: A unifying approach to the strong and weak measurement
  regime},\ }\href {https://doi.org/10.1103/PhysRevResearch.3.033123}
  {\bibfield  {journal} {\bibinfo  {journal} {Phys. Rev. Res.}\ }\textbf
  {\bibinfo {volume} {3}},\ \bibinfo {pages} {033123} (\bibinfo {year}
  {2021})}\BibitemShut {NoStop}%
\bibitem [{\citenamefont {Kerski}\ \emph {et~al.}(2023)\citenamefont {Kerski},
  \citenamefont {Mannel}, \citenamefont {Lochner}, \citenamefont
  {Kleinherbers}, \citenamefont {Kurzmann}, \citenamefont {Ludwig},
  \citenamefont {Wieck}, \citenamefont {K\"onig}, \citenamefont {Lorke},\ and\
  \citenamefont {Geller}}]{kerskiSR2023}%
  \BibitemOpen
  \bibfield  {author} {\bibinfo {author} {\bibfnamefont {J.}~\bibnamefont
  {Kerski}}, \bibinfo {author} {\bibfnamefont {H.}~\bibnamefont {Mannel}},
  \bibinfo {author} {\bibfnamefont {P.}~\bibnamefont {Lochner}}, \bibinfo
  {author} {\bibfnamefont {E.}~\bibnamefont {Kleinherbers}}, \bibinfo {author}
  {\bibfnamefont {A.}~\bibnamefont {Kurzmann}}, \bibinfo {author}
  {\bibfnamefont {A.}~\bibnamefont {Ludwig}}, \bibinfo {author} {\bibfnamefont
  {A.~D.}\ \bibnamefont {Wieck}}, \bibinfo {author} {\bibfnamefont
  {J.}~\bibnamefont {K\"onig}}, \bibinfo {author} {\bibfnamefont
  {A.}~\bibnamefont {Lorke}},\ and\ \bibinfo {author} {\bibfnamefont
  {M.}~\bibnamefont {Geller}},\ }\bibfield  {title} {\bibinfo {title}
  {Post-processing of real-time quantum event measurements for an optimal
  bandwidth},\ }\href {https://doi.org/10.1038/s41598-023-28273-0} {\bibfield
  {journal} {\bibinfo  {journal} {Sci. Rep.}\ }\textbf {\bibinfo {volume}
  {13}},\ \bibinfo {pages} {1105} (\bibinfo {year} {2023})}\BibitemShut
  {NoStop}%
\bibitem [{\citenamefont {H\"agele}\ and\ \citenamefont
  {Schefczik}(2018)}]{hagelePRB2018}%
  \BibitemOpen
  \bibfield  {author} {\bibinfo {author} {\bibfnamefont {D.}~\bibnamefont
  {H\"agele}}\ and\ \bibinfo {author} {\bibfnamefont {F.}~\bibnamefont
  {Schefczik}},\ }\bibfield  {title} {\bibinfo {title} {Higher-order moments,
  cumulants, and spectra of continuous quantum noise measurements},\
  }\href@noop {} {\bibfield  {journal} {\bibinfo  {journal} {Phys. Rev. B}\
  }\textbf {\bibinfo {volume} {98}},\ \bibinfo {pages} {205143} (\bibinfo
  {year} {2018})}\BibitemShut {NoStop}%
\bibitem [{\citenamefont {H\"agele}\ \emph {et~al.}(2020)\citenamefont
  {H\"agele}, \citenamefont {Sifft},\ and\ \citenamefont
  {Schefczik}}]{hagelePRB2020E}%
  \BibitemOpen
  \bibfield  {author} {\bibinfo {author} {\bibfnamefont {D.}~\bibnamefont
  {H\"agele}}, \bibinfo {author} {\bibfnamefont {M.}~\bibnamefont {Sifft}},\
  and\ \bibinfo {author} {\bibfnamefont {F.}~\bibnamefont {Schefczik}},\
  }\bibfield  {title} {\bibinfo {title} {Erratum: {Higher-order} moments,
  cumulants, and spectra of continuous quantum noise measurements { [Phys. Rev.
  B 98, 205143 (2018)]}},\ }\href@noop {} {\bibfield  {journal} {\bibinfo
  {journal} {Phys. Rev. B}\ }\textbf {\bibinfo {volume} {102}},\ \bibinfo
  {pages} {119901} (\bibinfo {year} {2020})}\BibitemShut {NoStop}%
\bibitem [{\citenamefont {Sifft}\ and\ \citenamefont
  {H\"agele}(2023)}]{sifftPRA2023}%
  \BibitemOpen
  \bibfield  {author} {\bibinfo {author} {\bibfnamefont {M.}~\bibnamefont
  {Sifft}}\ and\ \bibinfo {author} {\bibfnamefont {D.}~\bibnamefont
  {H\"agele}},\ }\bibfield  {title} {\bibinfo {title} {Random-time quantum
  measurements},\ }\href {https://doi.org/10.1103/PhysRevA.107.052203}
  {\bibfield  {journal} {\bibinfo  {journal} {Phys. Rev. A}\ }\textbf {\bibinfo
  {volume} {107}},\ \bibinfo {pages} {052203} (\bibinfo {year}
  {2023})}\BibitemShut {NoStop}%
\bibitem [{\citenamefont {Meinel}\ \emph {et~al.}(2022)\citenamefont {Meinel},
  \citenamefont {Vorobyov}, \citenamefont {Wang}, \citenamefont {Yavkin},
  \citenamefont {Pfender}, \citenamefont {Sumiya}, \citenamefont {Onada},
  \citenamefont {Isoya}, \citenamefont {Liu},\ and\ \citenamefont
  {Wrachtrup}}]{meinelNATURECOMM2022}%
  \BibitemOpen
  \bibfield  {author} {\bibinfo {author} {\bibfnamefont {J.}~\bibnamefont
  {Meinel}}, \bibinfo {author} {\bibfnamefont {V.}~\bibnamefont {Vorobyov}},
  \bibinfo {author} {\bibfnamefont {P.}~\bibnamefont {Wang}}, \bibinfo {author}
  {\bibfnamefont {B.}~\bibnamefont {Yavkin}}, \bibinfo {author} {\bibfnamefont
  {M.}~\bibnamefont {Pfender}}, \bibinfo {author} {\bibfnamefont
  {H.}~\bibnamefont {Sumiya}}, \bibinfo {author} {\bibfnamefont
  {S.}~\bibnamefont {Onada}}, \bibinfo {author} {\bibfnamefont
  {J.}~\bibnamefont {Isoya}}, \bibinfo {author} {\bibfnamefont {R.-B.}\
  \bibnamefont {Liu}},\ and\ \bibinfo {author} {\bibfnamefont {J.}~\bibnamefont
  {Wrachtrup}},\ }\bibfield  {title} {\bibinfo {title} {Quantum nonlinear
  spectroscopy of single nuclear spins},\ }\href@noop {} {\bibfield  {journal}
  {\bibinfo  {journal} {Nature Commun.}\ }\textbf {\bibinfo {volume} {13}},\
  \bibinfo {pages} {5318} (\bibinfo {year} {2022})}\BibitemShut {NoStop}%
\bibitem [{\citenamefont {Kleinherbers}\ \emph {et~al.}(2022)\citenamefont
  {Kleinherbers}, \citenamefont {Stegmann}, \citenamefont {Kurzmann},
  \citenamefont {Geller}, \citenamefont {Lorke},\ and\ \citenamefont
  {K\"onig}}]{kleinherbersPRL2022}%
  \BibitemOpen
  \bibfield  {author} {\bibinfo {author} {\bibfnamefont {E.}~\bibnamefont
  {Kleinherbers}}, \bibinfo {author} {\bibfnamefont {P.}~\bibnamefont
  {Stegmann}}, \bibinfo {author} {\bibfnamefont {A.}~\bibnamefont {Kurzmann}},
  \bibinfo {author} {\bibfnamefont {M.}~\bibnamefont {Geller}}, \bibinfo
  {author} {\bibfnamefont {A.}~\bibnamefont {Lorke}},\ and\ \bibinfo {author}
  {\bibfnamefont {J.}~\bibnamefont {K\"onig}},\ }\bibfield  {title} {\bibinfo
  {title} {Pushing the limits in real-time measurements of quantum dynamics},\
  }\href@noop {} {\bibfield  {journal} {\bibinfo  {journal} {Phys. Rev. Lett.}\
  }\textbf {\bibinfo {volume} {128}},\ \bibinfo {pages} {087701} (\bibinfo
  {year} {2022})}\BibitemShut {NoStop}%
\bibitem [{lan(2023)}]{landiARXIV2023}%
  \BibitemOpen
  \href@noop {} {} (\bibinfo {year} {2023}),\ \bibinfo {note} {{Gabriel T.
  Landi, Michael J. Kewming, Mark T. Mitchison, Patrick P. Potts}, Current
  fluctuations in open quantum systems: Bridging the gap between quantum
  continuous measurements and full counting statistics, arXiv:2303.04270
  [quant-ph]}\BibitemShut {NoStop}%
\bibitem [{\citenamefont {Cox}(1955)}]{coxJRSSSB1955}%
  \BibitemOpen
  \bibfield  {author} {\bibinfo {author} {\bibfnamefont {D.~R.}\ \bibnamefont
  {Cox}},\ }\bibfield  {title} {\bibinfo {title} {Some statistical methods
  connected with series of events},\ }\href
  {https://doi.org/https://doi.org/10.1111/j.2517-6161.1955.tb00188.x}
  {\bibfield  {journal} {\bibinfo  {journal} {J. R. Stat. Soc. Ser. B}\
  }\textbf {\bibinfo {volume} {17}},\ \bibinfo {pages} {129} (\bibinfo {year}
  {1955})}\BibitemShut {NoStop}%
\bibitem [{Kur()}]{KurzmannSupp2019}%
  \BibitemOpen
  \href@noop {} {}\bibinfo {note} {Supplemental Material of Ref.
  \onlinecite{kurzmannPRL2019}.}\BibitemShut {Stop}%
\bibitem [{\citenamefont {Brillinger}(1965)}]{brillingerAMS1965}%
  \BibitemOpen
  \bibfield  {author} {\bibinfo {author} {\bibfnamefont {D.~R.}\ \bibnamefont
  {Brillinger}},\ }\bibfield  {title} {\bibinfo {title} {An introduction to
  polyspectra},\ }\href@noop {} {\bibfield  {journal} {\bibinfo  {journal}
  {Ann. Math. Statist.}\ }\textbf {\bibinfo {volume} {36}},\ \bibinfo {pages}
  {1351} (\bibinfo {year} {1965})}\BibitemShut {NoStop}%
\bibitem [{\citenamefont {Gardiner}(2009)}]{gardinerBOOK2009}%
  \BibitemOpen
  \bibfield  {author} {\bibinfo {author} {\bibfnamefont {C.}~\bibnamefont
  {Gardiner}},\ }\href@noop {} {\emph {\bibinfo {title} {Stochastic
  Methods}}},\ \bibinfo {edition} {4th}\ ed.\ (\bibinfo  {publisher}
  {Springer},\ \bibinfo {address} {Berlin Heidelberg},\ \bibinfo {year}
  {2009})\BibitemShut {NoStop}%
\bibitem [{\citenamefont {Starosielec}\ \emph {et~al.}(2010)\citenamefont
  {Starosielec}, \citenamefont {Fainblat}, \citenamefont {Rudolph},\ and\
  \citenamefont {H\"agele}}]{starosielecRSI2010}%
  \BibitemOpen
  \bibfield  {author} {\bibinfo {author} {\bibfnamefont {S.}~\bibnamefont
  {Starosielec}}, \bibinfo {author} {\bibfnamefont {R.}~\bibnamefont
  {Fainblat}}, \bibinfo {author} {\bibfnamefont {J.}~\bibnamefont {Rudolph}},\
  and\ \bibinfo {author} {\bibfnamefont {D.}~\bibnamefont {H\"agele}},\
  }\bibfield  {title} {\bibinfo {title} {Two-dimensional higher order noise
  spectroscopy up to radio frequencies},\ }\href@noop {} {\bibfield  {journal}
  {\bibinfo  {journal} {Rev. Scientific Instrum.}\ }\textbf {\bibinfo {volume}
  {81}},\ \bibinfo {pages} {125101} (\bibinfo {year} {2010})}\BibitemShut
  {NoStop}%
\bibitem [{sif(2022)}]{sifftSIGNALSNAP2022}%
  \BibitemOpen
  \href@noop {} {} (\bibinfo {year} {2022}),\ \bibinfo {note} {{M. Sifft,
  SignalSnap Toolbox, https://github.com/MarkusSifft/SignalSnap}}\BibitemShut
  {NoStop}%
\bibitem [{\citenamefont {Forney}(1973)}]{forneyIEEE1973}%
  \BibitemOpen
  \bibfield  {author} {\bibinfo {author} {\bibfnamefont {G.~D.}\ \bibnamefont
  {Forney}},\ }\bibfield  {title} {\bibinfo {title} {The {Viterbi} algorithm},\
  }\href@noop {} {\bibfield  {journal} {\bibinfo  {journal} {Proc. IEEE}\
  }\textbf {\bibinfo {volume} {61}},\ \bibinfo {pages} {268} (\bibinfo {year}
  {1973})}\BibitemShut {NoStop}%
\bibitem [{\citenamefont {Rabiner}\ and\ \citenamefont
  {Juang}(1986)}]{rabinerIEEE1986}%
  \BibitemOpen
  \bibfield  {author} {\bibinfo {author} {\bibfnamefont {L.~R.}\ \bibnamefont
  {Rabiner}}\ and\ \bibinfo {author} {\bibfnamefont {B.~H.}\ \bibnamefont
  {Juang}},\ }\bibfield  {title} {\bibinfo {title} {An introduction to hidden
  {Markov} models},\ }\href {https://doi.org/10.1109/MASSP.1986.1165342}
  {\bibfield  {journal} {\bibinfo  {journal} {IEEE ASSP}\ }\textbf {\bibinfo
  {volume} {3}},\ \bibinfo {pages} {4} (\bibinfo {year} {1986})}\BibitemShut
  {NoStop}%
\bibitem [{\citenamefont {Bagrets}\ and\ \citenamefont
  {Nazarov}(2003)}]{bagretsPRB2003}%
  \BibitemOpen
  \bibfield  {author} {\bibinfo {author} {\bibfnamefont {D.~A.}\ \bibnamefont
  {Bagrets}}\ and\ \bibinfo {author} {\bibfnamefont {Y.~V.}\ \bibnamefont
  {Nazarov}},\ }\bibfield  {title} {\bibinfo {title} {Full counting statistics
  of charge transfer in {Coulomb} blockade systems},\ }\href@noop {} {\bibfield
   {journal} {\bibinfo  {journal} {Phys. Rev. B}\ }\textbf {\bibinfo {volume}
  {67}},\ \bibinfo {pages} {085316} (\bibinfo {year} {2003})}\BibitemShut
  {NoStop}%
\bibitem [{\citenamefont {Ubbelohde}\ \emph {et~al.}(2012)\citenamefont
  {Ubbelohde}, \citenamefont {Fricke}, \citenamefont {Flindt}, \citenamefont
  {Hohls},\ and\ \citenamefont {Haug}}]{ubbelohdeNATCOMM2012}%
  \BibitemOpen
  \bibfield  {author} {\bibinfo {author} {\bibfnamefont {N.}~\bibnamefont
  {Ubbelohde}}, \bibinfo {author} {\bibfnamefont {C.}~\bibnamefont {Fricke}},
  \bibinfo {author} {\bibfnamefont {C.}~\bibnamefont {Flindt}}, \bibinfo
  {author} {\bibfnamefont {F.}~\bibnamefont {Hohls}},\ and\ \bibinfo {author}
  {\bibfnamefont {R.~J.}\ \bibnamefont {Haug}},\ }\bibfield  {title} {\bibinfo
  {title} {Measurement of finite-frequency current statistics in a
  single-electron transistor},\ }\href {https://doi.org/10.1038/ncomms1620}
  {\bibfield  {journal} {\bibinfo  {journal} {Nat. Commun.}\ }\textbf {\bibinfo
  {volume} {3}},\ \bibinfo {pages} {612} (\bibinfo {year} {2012})}\BibitemShut
  {NoStop}%
\bibitem [{\citenamefont {Emary}\ \emph {et~al.}(2007)\citenamefont {Emary},
  \citenamefont {Marcos}, \citenamefont {Aguado},\ and\ \citenamefont
  {Brandes}}]{emaryPRB2007}%
  \BibitemOpen
  \bibfield  {author} {\bibinfo {author} {\bibfnamefont {C.}~\bibnamefont
  {Emary}}, \bibinfo {author} {\bibfnamefont {D.}~\bibnamefont {Marcos}},
  \bibinfo {author} {\bibfnamefont {R.}~\bibnamefont {Aguado}},\ and\ \bibinfo
  {author} {\bibfnamefont {T.}~\bibnamefont {Brandes}},\ }\bibfield  {title}
  {\bibinfo {title} {Frequency-dependent counting statistics in interacting
  nanoscale conductors},\ }\href@noop {} {\bibfield  {journal} {\bibinfo
  {journal} {Phys. Rev. B}\ }\textbf {\bibinfo {volume} {76}},\ \bibinfo
  {pages} {161404} (\bibinfo {year} {2007})}\BibitemShut {NoStop}%
\bibitem [{\citenamefont {Kambly}\ \emph {et~al.}(2011)\citenamefont {Kambly},
  \citenamefont {Flindt},\ and\ \citenamefont {B\"uttiker}}]{kamblyPRB2011}%
  \BibitemOpen
  \bibfield  {author} {\bibinfo {author} {\bibfnamefont {D.}~\bibnamefont
  {Kambly}}, \bibinfo {author} {\bibfnamefont {C.}~\bibnamefont {Flindt}},\
  and\ \bibinfo {author} {\bibfnamefont {M.}~\bibnamefont {B\"uttiker}},\
  }\bibfield  {title} {\bibinfo {title} {Factorial cumulants reveal
  interactions in counting statistics},\ }\href
  {https://doi.org/10.1103/PhysRevB.83.075432} {\bibfield  {journal} {\bibinfo
  {journal} {Phys. Rev. B}\ }\textbf {\bibinfo {volume} {83}},\ \bibinfo
  {pages} {075432} (\bibinfo {year} {2011})}\BibitemShut {NoStop}%
\bibitem [{\citenamefont {Stegmann}\ \emph {et~al.}(2015)\citenamefont
  {Stegmann}, \citenamefont {Sothmann}, \citenamefont {Hucht},\ and\
  \citenamefont {K\"onig}}]{stegmannPRB2015}%
  \BibitemOpen
  \bibfield  {author} {\bibinfo {author} {\bibfnamefont {P.}~\bibnamefont
  {Stegmann}}, \bibinfo {author} {\bibfnamefont {B.}~\bibnamefont {Sothmann}},
  \bibinfo {author} {\bibfnamefont {A.}~\bibnamefont {Hucht}},\ and\ \bibinfo
  {author} {\bibfnamefont {J.}~\bibnamefont {K\"onig}},\ }\bibfield  {title}
  {\bibinfo {title} {Detection of interactions via generalized factorial
  cumulants in systems in and out of equilibrium},\ }\href
  {https://doi.org/10.1103/PhysRevB.92.155413} {\bibfield  {journal} {\bibinfo
  {journal} {Phys. Rev. B}\ }\textbf {\bibinfo {volume} {92}},\ \bibinfo
  {pages} {155413} (\bibinfo {year} {2015})}\BibitemShut {NoStop}%
\bibitem [{\citenamefont {Tilloy}(2018)}]{tilloyPRA2018}%
  \BibitemOpen
  \bibfield  {author} {\bibinfo {author} {\bibfnamefont {A.}~\bibnamefont
  {Tilloy}},\ }\bibfield  {title} {\bibinfo {title} {Exact signal correlators
  in continuous quantum measurements},\ }\href@noop {} {\bibfield  {journal}
  {\bibinfo  {journal} {Phys. Rev. A}\ }\textbf {\bibinfo {volume} {98}},\
  \bibinfo {pages} {010104(R)} (\bibinfo {year} {2018})}\BibitemShut {NoStop}%
\bibitem [{\citenamefont {Jacobs}\ and\ \citenamefont
  {Steck}(2006)}]{jacobsCP2006}%
  \BibitemOpen
  \bibfield  {author} {\bibinfo {author} {\bibfnamefont {K.}~\bibnamefont
  {Jacobs}}\ and\ \bibinfo {author} {\bibfnamefont {D.~A.}\ \bibnamefont
  {Steck}},\ }\bibfield  {title} {\bibinfo {title} {A straightforward
  introduction to continuous quantum measurement},\ }\href
  {https://doi.org/10.1080/00107510601101934} {\bibfield  {journal} {\bibinfo
  {journal} {Contemp. Phys.}\ }\textbf {\bibinfo {volume} {47}},\ \bibinfo
  {pages} {279} (\bibinfo {year} {2006})}\BibitemShut {NoStop}%
\bibitem [{\citenamefont {Barchielli}\ \emph {et~al.}(1982)\citenamefont
  {Barchielli}, \citenamefont {Lanz},\ and\ \citenamefont
  {Prosperi}}]{barchielliNC1982}%
  \BibitemOpen
  \bibfield  {author} {\bibinfo {author} {\bibfnamefont {A.}~\bibnamefont
  {Barchielli}}, \bibinfo {author} {\bibfnamefont {L.}~\bibnamefont {Lanz}},\
  and\ \bibinfo {author} {\bibfnamefont {G.~M.}\ \bibnamefont {Prosperi}},\
  }\bibfield  {title} {\bibinfo {title} {A model for the macroscopic
  description and continual observations in quantum mechanics},\ }\href@noop {}
  {\bibfield  {journal} {\bibinfo  {journal} {Nuovo Cimento}\ }\textbf
  {\bibinfo {volume} {72B}},\ \bibinfo {pages} {79} (\bibinfo {year}
  {1982})}\BibitemShut {NoStop}%
\bibitem [{\citenamefont {Barchielli}\ and\ \citenamefont
  {Gregoratti}(2009)}]{barchielliBOOK2009}%
  \BibitemOpen
  \bibfield  {author} {\bibinfo {author} {\bibfnamefont {A.}~\bibnamefont
  {Barchielli}}\ and\ \bibinfo {author} {\bibfnamefont {M.}~\bibnamefont
  {Gregoratti}},\ }\href@noop {} {\emph {\bibinfo {title} {Quantum Trajectories
  and Measurements in Continuous Time: The Diffusive Case}}},\ Lecture Notes in
  Physics 782\ (\bibinfo  {publisher} {Springer},\ \bibinfo {address} {Berlin
  Heidelberg},\ \bibinfo {year} {2009})\BibitemShut {NoStop}%
\bibitem [{\citenamefont {Belavkin}(1987)}]{belavkinConf1987}%
  \BibitemOpen
  \bibfield  {author} {\bibinfo {author} {\bibfnamefont {V.}~\bibnamefont
  {Belavkin}},\ }\bibfield  {title} {\bibinfo {title} {Non-demolition
  measurement and control in quantum dynamical systems},\ }in\ \href@noop {}
  {\emph {\bibinfo {booktitle} {Information Complexity and Control in Quantum
  Physics}}},\ \bibinfo {series} {International Centre for Mechanical
  Sciences}, Vol.\ \bibinfo {volume} {294},\ \bibinfo {editor} {edited by\
  \bibinfo {editor} {\bibfnamefont {A.}~\bibnamefont {Blaquiere}}, \bibinfo
  {editor} {\bibfnamefont {S.}~\bibnamefont {Diner}},\ and\ \bibinfo {editor}
  {\bibfnamefont {G.}~\bibnamefont {Lochak}}}\ (\bibinfo {address} {Vienna},\
  \bibinfo {year} {1987})\ p.\ \bibinfo {pages} {311}\BibitemShut {NoStop}%
\bibitem [{\citenamefont {Diosi}(1988)}]{diosiPLA1988}%
  \BibitemOpen
  \bibfield  {author} {\bibinfo {author} {\bibfnamefont {L.}~\bibnamefont
  {Diosi}},\ }\bibfield  {title} {\bibinfo {title} {Continuous quantum
  measurement and {Ito} formalism},\ }\href@noop {} {\bibfield  {journal}
  {\bibinfo  {journal} {Phys. Lett. A}\ }\textbf {\bibinfo {volume} {129}},\
  \bibinfo {pages} {419} (\bibinfo {year} {1988})}\BibitemShut {NoStop}%
\bibitem [{\citenamefont {Gagen}\ \emph {et~al.}(1993)\citenamefont {Gagen},
  \citenamefont {Wiseman},\ and\ \citenamefont {Milburn}}]{gagenPRA1993}%
  \BibitemOpen
  \bibfield  {author} {\bibinfo {author} {\bibfnamefont {M.~J.}\ \bibnamefont
  {Gagen}}, \bibinfo {author} {\bibfnamefont {H.~M.}\ \bibnamefont {Wiseman}},\
  and\ \bibinfo {author} {\bibfnamefont {G.~J.}\ \bibnamefont {Milburn}},\
  }\bibfield  {title} {\bibinfo {title} {Continuous position measurements and
  the quantum {Zeno} effect},\ }\href@noop {} {\bibfield  {journal} {\bibinfo
  {journal} {Phys. Rev. A}\ }\textbf {\bibinfo {volume} {48}},\ \bibinfo
  {pages} {132} (\bibinfo {year} {1993})}\BibitemShut {NoStop}%
\bibitem [{\citenamefont {Korotkov}(1999)}]{korotkovPRB1999}%
  \BibitemOpen
  \bibfield  {author} {\bibinfo {author} {\bibfnamefont {A.~N.}\ \bibnamefont
  {Korotkov}},\ }\bibfield  {title} {\bibinfo {title} {Continuous quantum
  measurement of a double dot},\ }\href@noop {} {\bibfield  {journal} {\bibinfo
   {journal} {Phys. Rev. B}\ }\textbf {\bibinfo {volume} {60}},\ \bibinfo
  {pages} {5737} (\bibinfo {year} {1999})}\BibitemShut {NoStop}%
\bibitem [{\citenamefont {Korotkov}(2001{\natexlab{a}})}]{korotkovPRB2001}%
  \BibitemOpen
  \bibfield  {author} {\bibinfo {author} {\bibfnamefont {A.~N.}\ \bibnamefont
  {Korotkov}},\ }\bibfield  {title} {\bibinfo {title} {Output spectrum of a
  detector measuring quantum oscillations},\ }\href@noop {} {\bibfield
  {journal} {\bibinfo  {journal} {Phys. Rev. B}\ }\textbf {\bibinfo {volume}
  {63}},\ \bibinfo {pages} {085312} (\bibinfo {year}
  {2001}{\natexlab{a}})}\BibitemShut {NoStop}%
\bibitem [{\citenamefont {Goan}\ \emph {et~al.}(2001)\citenamefont {Goan},
  \citenamefont {Milburn}, \citenamefont {Wiseman},\ and\ \citenamefont
  {Sun}}]{goanPRB2001}%
  \BibitemOpen
  \bibfield  {author} {\bibinfo {author} {\bibfnamefont {H.-S.}\ \bibnamefont
  {Goan}}, \bibinfo {author} {\bibfnamefont {G.~J.}\ \bibnamefont {Milburn}},
  \bibinfo {author} {\bibfnamefont {H.~M.}\ \bibnamefont {Wiseman}},\ and\
  \bibinfo {author} {\bibfnamefont {H.~B.}\ \bibnamefont {Sun}},\ }\bibfield
  {title} {\bibinfo {title} {Continuous quantum measurement of two coupled
  quantum dots using a point contact: A quantum trajectory approach},\ }\href
  {https://doi.org/10.1103/PhysRevB.63.125326} {\bibfield  {journal} {\bibinfo
  {journal} {Phys. Rev. B}\ }\textbf {\bibinfo {volume} {63}},\ \bibinfo
  {pages} {125326} (\bibinfo {year} {2001})}\BibitemShut {NoStop}%
\bibitem [{\citenamefont {Attal}\ and\ \citenamefont
  {Pautrat}(2006)}]{Attal2006}%
  \BibitemOpen
  \bibfield  {author} {\bibinfo {author} {\bibfnamefont {S.}~\bibnamefont
  {Attal}}\ and\ \bibinfo {author} {\bibfnamefont {Y.}~\bibnamefont
  {Pautrat}},\ }\bibfield  {title} {\bibinfo {title} {From repeated to
  continuous quantum interactions},\ }\href
  {https://doi.org/10.1007/s00023-005-0242-8} {\bibfield  {journal} {\bibinfo
  {journal} {Annales Henri Poincare}\ }\textbf {\bibinfo {volume} {7}},\
  \bibinfo {pages} {59} (\bibinfo {year} {2006})}\BibitemShut {NoStop}%
\bibitem [{\citenamefont {Attal}\ and\ \citenamefont
  {Pellegrini}(2010)}]{Attal2010}%
  \BibitemOpen
  \bibfield  {author} {\bibinfo {author} {\bibfnamefont {S.}~\bibnamefont
  {Attal}}\ and\ \bibinfo {author} {\bibfnamefont {C.}~\bibnamefont
  {Pellegrini}},\ }\bibfield  {title} {\bibinfo {title} {Stochastic master
  equations in thermal environment},\ }\href
  {https://doi.org/10.1142/S1230161210000242} {\bibfield  {journal} {\bibinfo
  {journal} {Open Syst. Inf. Dyn.}\ }\textbf {\bibinfo {volume} {17}},\
  \bibinfo {pages} {389} (\bibinfo {year} {2010})}\BibitemShut {NoStop}%
\bibitem [{\citenamefont {Gross}\ \emph {et~al.}(2018)\citenamefont {Gross},
  \citenamefont {Caves}, \citenamefont {Milburn},\ and\ \citenamefont
  {Combes}}]{Gross2018}%
  \BibitemOpen
  \bibfield  {author} {\bibinfo {author} {\bibfnamefont {J.}~\bibnamefont
  {Gross}}, \bibinfo {author} {\bibfnamefont {C.}~\bibnamefont {Caves}},
  \bibinfo {author} {\bibfnamefont {G.}~\bibnamefont {Milburn}},\ and\ \bibinfo
  {author} {\bibfnamefont {J.}~\bibnamefont {Combes}},\ }\bibfield  {title}
  {\bibinfo {title} {Qubit models of weak continuous measurements: Markovian
  conditional and open-system dynamics},\ }\href@noop {} {\bibfield  {journal}
  {\bibinfo  {journal} {Quantum Science and Technology}\ }\textbf {\bibinfo
  {volume} {3}},\ \bibinfo {pages} {024005} (\bibinfo {year}
  {2018})}\BibitemShut {NoStop}%
\bibitem [{\citenamefont
  {Korotkov}(2001{\natexlab{b}})}]{korotkovPhysRevB2001}%
  \BibitemOpen
  \bibfield  {author} {\bibinfo {author} {\bibfnamefont {A.~N.}\ \bibnamefont
  {Korotkov}},\ }\bibfield  {title} {\bibinfo {title} {Output spectrum of a
  detector measuring quantum oscillations},\ }\href
  {https://doi.org/10.1103/PhysRevB.63.085312} {\bibfield  {journal} {\bibinfo
  {journal} {Phys. Rev. B}\ }\textbf {\bibinfo {volume} {63}},\ \bibinfo
  {pages} {085312} (\bibinfo {year} {2001}{\natexlab{b}})}\BibitemShut
  {NoStop}%
\bibitem [{\citenamefont {Johansson}\ \emph {et~al.}(2013)\citenamefont
  {Johansson}, \citenamefont {Nation},\ and\ \citenamefont
  {Nori}}]{JOHANSSON20131234}%
  \BibitemOpen
  \bibfield  {author} {\bibinfo {author} {\bibfnamefont {J.}~\bibnamefont
  {Johansson}}, \bibinfo {author} {\bibfnamefont {P.}~\bibnamefont {Nation}},\
  and\ \bibinfo {author} {\bibfnamefont {F.}~\bibnamefont {Nori}},\ }\bibfield
  {title} {\bibinfo {title} {Qutip 2: A {Python} framework for the dynamics of
  open quantum systems},\ }\href
  {https://doi.org/https://doi.org/10.1016/j.cpc.2012.11.019} {\bibfield
  {journal} {\bibinfo  {journal} {Comput. Phys. Commun.}\ }\textbf {\bibinfo
  {volume} {184}},\ \bibinfo {pages} {1234 } (\bibinfo {year}
  {2013})}\BibitemShut {NoStop}%
\bibitem [{\citenamefont {Tilloy}\ \emph {et~al.}(2015)\citenamefont {Tilloy},
  \citenamefont {Bauer},\ and\ \citenamefont {Bernard}}]{TilloyPRA2015}%
  \BibitemOpen
  \bibfield  {author} {\bibinfo {author} {\bibfnamefont {A.}~\bibnamefont
  {Tilloy}}, \bibinfo {author} {\bibfnamefont {M.}~\bibnamefont {Bauer}},\ and\
  \bibinfo {author} {\bibfnamefont {D.}~\bibnamefont {Bernard}},\ }\bibfield
  {title} {\bibinfo {title} {Spikes in quantum trajectories},\ }\href
  {https://doi.org/10.1103/PhysRevA.92.052111} {\bibfield  {journal} {\bibinfo
  {journal} {Phys. Rev. A}\ }\textbf {\bibinfo {volume} {92}},\ \bibinfo
  {pages} {052111} (\bibinfo {year} {2015})}\BibitemShut {NoStop}%
\bibitem [{foo()}]{footnote1}%
  \BibitemOpen
  \href@noop {} {\bibinfo {title} {{Please note, that $k$ is not required on
  the right-hand side.}}}\BibitemShut {Stop}%
\bibitem [{sif(2024)}]{sifftQuantumCatch2024}%
  \BibitemOpen
  \href@noop {} {} (\bibinfo {year} {2024}),\ \bibinfo {note} {{M. Sifft,
  QuantumCatch Toolbox,
  https://github.com/MarkusSifft/QuantumCatch}}\BibitemShut {NoStop}%
\bibitem [{sch(2019)}]{schefczikARXIV2019}%
  \BibitemOpen
  \href@noop {} {} (\bibinfo {year} {2019}),\ \bibinfo {note} {{F. Schefczik
  and D. H\"agele}, arXiv:1904.12154v1 [math.ST]}\BibitemShut {NoStop}%
\bibitem [{\citenamefont {Starosielec}\ and\ \citenamefont
  {H\"agele}(2014)}]{starosielecSP2014}%
  \BibitemOpen
  \bibfield  {author} {\bibinfo {author} {\bibfnamefont {S.}~\bibnamefont
  {Starosielec}}\ and\ \bibinfo {author} {\bibfnamefont {D.}~\bibnamefont
  {H\"agele}},\ }\bibfield  {title} {\bibinfo {title} {Discrete-time windows
  with minimal {RMS} bandwidth for given {RMS} temporal width},\ }\href@noop {}
  {\bibfield  {journal} {\bibinfo  {journal} {Signal Processing (Elsevier)}\
  }\textbf {\bibinfo {volume} {102}},\ \bibinfo {pages} {240} (\bibinfo {year}
  {2014})}\BibitemShut {NoStop}%
\bibitem [{\citenamefont {Fisher}(1928)}]{fisherPLMS1930}%
  \BibitemOpen
  \bibfield  {author} {\bibinfo {author} {\bibfnamefont {R.~A.}\ \bibnamefont
  {Fisher}},\ }\bibfield  {title} {\bibinfo {title} {Moments and product
  moments of sampling distributions},\ }\href@noop {} {\bibfield  {journal}
  {\bibinfo  {journal} {Proc. Lond. Math. Soc. 2nd Ser.}\ }\textbf {\bibinfo
  {volume} {30}},\ \bibinfo {pages} {199} (\bibinfo {year} {1928})}\BibitemShut
  {NoStop}%
\bibitem [{\citenamefont {Yalamanchili}\ \emph {et~al.}(2015)\citenamefont
  {Yalamanchili}, \citenamefont {Arshad}, \citenamefont {Mohammed},
  \citenamefont {Garigipati}, \citenamefont {Entschev}, \citenamefont
  {Kloppenborg}, \citenamefont {Malcolm},\ and\ \citenamefont
  {Melonakos}}]{Yalamanchili2015}%
  \BibitemOpen
  \bibfield  {author} {\bibinfo {author} {\bibfnamefont {P.}~\bibnamefont
  {Yalamanchili}}, \bibinfo {author} {\bibfnamefont {U.}~\bibnamefont
  {Arshad}}, \bibinfo {author} {\bibfnamefont {Z.}~\bibnamefont {Mohammed}},
  \bibinfo {author} {\bibfnamefont {P.}~\bibnamefont {Garigipati}}, \bibinfo
  {author} {\bibfnamefont {P.}~\bibnamefont {Entschev}}, \bibinfo {author}
  {\bibfnamefont {B.}~\bibnamefont {Kloppenborg}}, \bibinfo {author}
  {\bibfnamefont {J.}~\bibnamefont {Malcolm}},\ and\ \bibinfo {author}
  {\bibfnamefont {J.}~\bibnamefont {Melonakos}},\ }\href
  {https://github.com/arrayfire/arrayfire} {\bibinfo {title} {{ArrayFire - A
  high performance software library for parallel computing with an easy-to-use
  API}}} (\bibinfo {year} {2015})\BibitemShut {NoStop}%
\end{thebibliography}%
\end{document}